%% file: TranIT11toArxiv.tex
\documentclass[journal]{IEEEtran}
\input{pream}
\usepackage{graphicx}
\usepackage{cite}
\usepackage{balance}
\usepackage{latexsym}
\newcounter{mytempeqncnt}
\allowdisplaybreaks
\ifCLASSINFOpdf
\else
\fi
\begin{document}
%
\title{Continuum Limits of Markov Chains with Application to Network Modeling}
%
%
%

\author{Yang~Zhang,~\IEEEmembership{Student Member,~IEEE,} Edwin~K. P.~Chong,~\IEEEmembership{Fellow,~IEEE,} Jan~Hannig,~
and~Donald~Estep
\thanks{This research was supported in part by NSF grant ECCS-0700559.}
\thanks{Yang Zhang and Edwin K. P. Chong are with the Department of Electrical and Computer Engineering, Colorado State University, Ft. Collins, CO 80523-1373
{\tt\small yzhangcn@mail.engr.colostate.edu \& edwin.chong@colostate.edu}}%
\thanks{Jan Hannig is with the Department of Statistics and Operation Research, The University of North Carolina at Chapel Hill, Chapel Hill, NC 27599-3260
{\tt\small jan.hannig@unc.edu}}%
\thanks{Donald Estep is with the Department of Mathematics and Department of Statistics, Colorado State University, Fort Collins, CO 80523-1373
{\tt\small estep@math.colostate.edu}}%
\thanks{A preliminary version of parts of the work of this paper was presented at the 49th IEEE Conference on Decision and Control.}
}
\maketitle

\begin{abstract}
In this paper we investigate the continuum limits of a class of
Markov chains. The investigation of such limits is motivated by the desire to model very large networks. We show that under some
conditions, a sequence of Markov chains converges in some sense to the solution of a partial differential equation. Based on such convergence we approximate Markov chains modeling networks with a large number of components by partial differential equations. While traditional Monte Carlo simulation for very large networks is practically infeasible, partial differential equations can be solved with reasonable computational overhead using well-established mathematical tools.
\end{abstract}

\begin{IEEEkeywords}
Continuum modeling, Markov chain, partial differential equation, large network modeling, wireless sensor network.
\end{IEEEkeywords}

%
\IEEEpeerreviewmaketitle
\section{Introduction}\label{sec:intro}
\IEEEPARstart{N}{etwork} modeling is an important tool in the analysis and design of networks. Many network characteristics of interest can be modeled by Markov chains, where Monte Carlo simulation has been the traditional approach~\cite{network_sim_book}. With the enormous growth in the size and complexity of today's networks, their simulation becomes more computationally expensive in both time and hardware. Some effort has been made to exploit the computing powers of distributed computer networks, such as parallel simulation techniques, where the number of processors needed in the simulation increases with the number of nodes in the network~\cite{paral_sim_1,paral_sim_2}.
However, for networks involving a very large number of nodes, Monte Carlo simulation eventually becomes practically infeasible.

In this paper we address this problem by focusing on the global characteristics of an entire network rather than those of its individual components.  The idea is to approximate the underlying Markov chain modeling a certain network characteristic by a partial differential equation (PDE).

As a concrete familiar example, which we present in Section~\ref{sec:randomwalk}, consider multiple i.i.d.\ (independent and identically distributed) random walks of $M$ particles on a network consisting of $N$ points. For any vector $x$, let $x^T$ be its transpose. Let the Markov chain modeling the network characteristic be $X_N(k)=[X_N(k,1),\ldots,X_N(k,N)]^T\in\real^N$, where $X_N(k,n)$ is the number of particles at point $n$ at time $k$. If we treat $N$ and $M$ as indices
that grow, this defines a family of Markov chains indexed by $N$ and $M$. We show that as $M \gt \infty$ and $N \gt \infty$, $X_N(k)$ converges in some sense to its continuum limit, a deterministic function with continuous time and space variables.
Under certain conditions, it is possible to characterize such a function as the solution of a PDE~\cite{kushnerbook,RePEc:spr:finsto:v:9:y:2005:i:4:p:519-537,darling-2008-5}. This itself is not a new result, but helps to illustrate our aim.

Indeed, our development here is motivated by the network modeling strategy in~\cite{chongcontinuum} and the need for a rigorous description of its underlying limiting process. We illustrate in Section~\ref{sec:convergence} the convergence of the sequence of Markov chains to the PDE in a two-step procedure.
Suppose the evolution of $X_N(k)$ is governed by a certain stochastic difference equation with a ``normalizing'' parameter $M$. Let $x_N(k)$ be the normalized deterministic sequence governed by the corresponding ``expected'' and deterministic difference equation. First, we show in Section~\ref{subsec:odeconvergence} that $X_N(k)/M$ is close to $x_N(k)$, in the sense that as $M\to\infty$, both their continuous-time extensions converge to the solution of an ordinary differential equation (ODE). Second, we show in Section~\ref{subsec:pdeconvergence} that as $N\to\infty$, $x_N(k)$ converges to the solution of a PDE. Therefore, as $M\to\infty$ and $N\to\infty$, $X_N(k)/M$ converges to the PDE solution.

Our procedure provides an approach to approximating Markov chains that model large networks by PDEs.
PDEs are widely used to formulate time-space phenomena in physics, chemistry, ecology, and economics (e.g., \cite{PDE1,PDE2,PDE3,PDE4}), and there are well-established mathematical tools for solving them such as Matlab and Comsol, which use finite element method~\cite{FEM} or finite difference method~\cite{FDM}.
In contrast to Monte Carlo simulation, our approach enables us to use these tools to greatly reduce computation time, which makes it possible to carry out the analysis, design, and
optimization for very large networks.
We present in Section~\ref{sec:networkexample} an example of the application of our approach to the modeling of a large wireless
sensor network. In this example, we derive an explicit nonlinear diffusion-convection PDE, whose solution captures the dynamic behavior of the data message queues in the network. We show that although the PDE approximation takes only a tiny fraction of the computation time of the Monte Carlo simulation, there is a strong agreement between their simulation results.

Continuum modeling has been well-established in fields such as physics, mechanics, transportation, and biology (e.g., \cite{cont_mdl_phy_book,cont_mdl_mtrl_book,cont_mdl_traf_book,cont_mdl_bio_paper}). Its applications in communication networks, however, are relatively new and rare. Among these, to our best knowledge, our approach is the first to address the time-space characteristics of communication networks with a large number of nodes. In contrast, for example, \cite{hvy_trffc_98,hvy_trffc_09,hvy_trffc_10} deal with networks with heavy traffic instead of large number of nodes; \cite{kumarnetworkcapacity,homo_cont_mdl} present scaling laws of the network traffic without characterizing the actual traffic over time and space; and~\cite{mean_fld_paper1,mean_fld_paper2}, which use mean field methods, only keep track of the statistical features of the networks such as the fraction of nodes in each network state.

\section{Continuum Limit of Multiple Random Walks} \label{sec:randomwalk}
\begin{figure}[htbp]
\centering
\includegraphics[width=70mm]{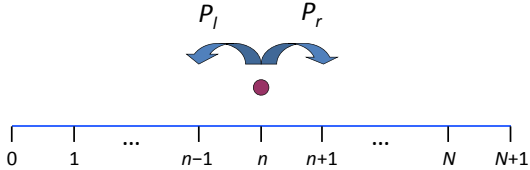}
\caption{An illustration of a one-dimensional single random walk.}
\label{fig:1DRandomWalk_conv}
\end{figure}

In this section we present an illustrative example of approximating multiple i.i.d.\ random walks by a PDE. First consider a single random walk on a one-dimensional network consisting of $N$ points uniformly placed over $\mathcal D=[0,1]$, as shown in Fig.~\ref{fig:1DRandomWalk_conv}. Hence the distance between two neighboring point is  $ds=1/(N+1)$. At each time instant, the particle at point $n$, where $n=1,\ldots,N$, randomly chooses to move to its left or right neighboring point with probability $P_r(n)$ and $P_l(n)$, respectively. Let the length between two time instants be $dt=1/M$. We set $dt=ds^2$, which is a standard time-space scaling approach to ensuring the convergence of the difference equation to a PDE.
We assume a ``sink'' boundary condition, i.e., the particle vanishes when it reaches the ends of $\mathcal D$ (though ``walls'' at the boundary are equally treatable).

Now consider $M$ random walks on the same network, where the particle in each random walk behaves independently identically as in the single random walk described above. Let $B_i(k,n)$ be the Bernoulli random variable representing the presence of the $i$th particle at point $n$ at time instant $k$, where $k=0,1,\ldots$ and $n=1,\ldots,N$.
Define $B_i(k)=[B_i(k,1),\ldots,B_i(k,N)]^T\in\real^N$.
According to the behavior of the particle in the single random walk, for $i=1,\ldots,M$,
\begin{align*}
&B_i(k+1,n)-B_i(k,n)\nn\\
&=\left\{
\begin{array}{lr}
B_i(k,n-1), &\mbox{ with probability } P_r(n-1);\\
B_i(k,n+1), &\mbox{ with probability } P_l(n+1);\\
-B_i(k,n), &\mbox{ with probability }P_r(n)+P_l(n);\\
0, &\mbox{ otherwise},\\
\end{array}
\right.
\end{align*}
where $B_i(k,n)$ with $n\leq0$ or $n\geq N+1$ are defined to be zero.
Let the function $F_N(x,U(k))$, where $U(k)$ are i.i.d.\ and do not depend on $x$, be such that for $i=1,\ldots,M$,
\beq
B_i(k+1) = B_i(k) + F_N(B_i(k),U(k)).\label{equ:rwB}
\eeq
Then for $x=[x_1,\ldots,x_N]^T$, the
$n$th component of $F_N(x,U(k))$, where $n=1,\ldots,N$, is
\begin{align}
&\left\{
\begin{array}{lr}
x_{n-1}, &\mbox{ with probability } P_r(n-1);\\
x_{n+1}, &\mbox{ with probability } P_l(n+1);\\
-x_n, &\mbox{ with probability  } P_r(n)+P_l(n);\\
0, &\mbox{ otherwise},
\end{array}\label{rwFN}
\right.
\end{align}
where $x_n$ with $n\leq0$ or $n\geq N+1$ are defined to be zero.

%
%

Let $X_N(k,n)$ be the number of particles at point $n$ at time $k$. Then
\beq
X_N(k,n) = \sum_{i=1}^{M}B_i(k,n).
\label{def:rwX}
\eeq
Define $X_N(k)=[X_N(k,1),\ldots,X_N(k,N)]^T$, which forms a discrete-time Markov chain with state space $\real ^N$.
Since $F_N$ is linear, it follows from~(\ref{def:rwX}) that
\beq
X_N(k+1) = X_N(k) + F_N(X_N(k),U(k)).\nn
\eeq

Let
\beq
f_N(x) = EF_N(x,U(k)),\quad x\in\real^N.\nn
\eeq
It follows from~(\ref{rwFN}) that for $x=[x_1,\ldots,x_N]^T$, the
$n$th component of $f_N(x)$, where $n=1,\ldots,N$, is
\beq
P_r(n-1)x_{n-1}+P_l(n+1)x_{n+1}-(P_r(n)+P_l(n))x_n,\label{rwfN}
\eeq
where $x_n$ with $n\leq0$ or $n\geq N+1$ are defined to be zero.
By~(\ref{equ:rwB}) and the linearity of $F_N$, for $i=1,\ldots,M$,
\beq
EB_i(k+1) = EB_i(k) + f_N(EB_i(k)).\label{equ:EB}
\eeq
Notice that, since the random walks are i.i.d., $EB_i$ does not depend on $i$.
Define a deterministic sequence $x_N(k)$ by
\beq
x_N(k+1) = x_N(k) + f_N(x_N(k)),\label{rwxfN}
\eeq
where
\beq
x_N(0)=\frac{X_N(0)}{M},\mbox{ a.s.\ (almost surely).}\label{equ:xX0}
\eeq

We seek to approximate $X_N(k)$ by a continuum model, where the time and space indices $k$ and $n$ are made continuous as $N\gt\infty$ and $M\gt\infty$ in the following two steps:
First, define
\[
X_{oN}(\tld t)=\frac{X_N(\floor{M\tld t})}{M},
\]
the continuous-time extension of $X_N(k)$ by piecewise-constant time extensions with interval length $dt=1/M$ and scaled by $1/M$. Second, define $X_{pN}(t,s)$ to be the continuous-space extension of $X_{oN}(\tld t)$ by piecewise-constant space extensions on $\mathcal D$ with interval length $ds$. Notice that as $N\gt\infty$, $ds\gt 0$. Thus $X_{pN}$ is the continuous-time-space extension of $X_N(k)$.
Similarly, define $x_{oN}(\tld t)=x_N(\floor{M\tld t})$, the piecewise-constant continuous-time extension of $x_N(k)$, and $x_{pN}(t,s)$, the piecewise-constant continuous-space extension of $x_{oN}(\tld t)$. Thus $x_{pN}$ is the continuous-time-space extension of $x_N(k)$.

Now we show that for $M$ sufficiently large, $X_{pN}$, the continuous-time-space extension of $X_N(k)$, is close to $x_{pN}$, the continuous-time-space extension of $x_N(k)$.
By~(\ref{def:rwX}) and the strong law of large numbers (SLLN), for each $k$,
\begin{align*}
&\lim_{M\gt\infty}\frac{X_N(k)}{M}= EB_i(k)\mbox{ a.s.}
\end{align*}
By this and~(\ref{equ:xX0}),
\[
\lim_{M\gt\infty}x_N(0)= EB_i(0)\mbox{ a.s.}
\]
By~(\ref{equ:EB}) and (\ref{rwxfN}), $x_N(k)$ and $EB_i(k)$ satisfy the same difference equation. Then  we have for each $k$,
\begin{align*}
\lim_{M\gt\infty}x_N(k)=EB_i(k)\mbox{ a.s.}
\end{align*}
Hence for each $k$,
\begin{align*}
&\lim_{M\gt\infty}\frac{X_N(k)}{M}= x_N(k) \mbox{ a.s.}
\end{align*}
Therefore, $X_{oN}$ and $x_{oN}$ are close for large $M$ in the sense that
\beq
\lim_{M\gt\infty}\|X_{oN}(\tld t)-x_{oN}(\tld t)\|_\infty^{(N)} = 0 \mbox{ a.s.},\label{equ:rwdo0}
\eeq
where $\|\cdot\|_\infty^{(N)}$ is the $\infty$-norm on $\real^N$.
Note that
\[
\|X_{pN}(\cdot,t)-x_{pN}(\cdot,t)\|_\infty^{(\mathcal D)}=\|X_{oN}-x_{oN}\|_\infty^{(N)},
\]
where $\|\cdot\|_\infty^{(\mathcal D)}$ is the $\infty$-norm on $\real^{\mathcal D}$, the space of functions of $\mathcal D\to\real$. Then by~(\ref{equ:rwdo0}), $X_{pN}$ and $x_{pN}$ are close to each other for large $M$ in the sense that
\beq
\lim_{M\gt\infty}\|X_{pN}(\cdot,t)-x_{pN}(\cdot,t)\|_\infty^{(\mathcal D)} = 0 \mbox{ a.s.}\label{equ:rwdp}
\eeq
Therefore, we can approximate $X_{pN}$ by $x_{pN}$ for $M$ sufficiently large.

Next we show that as $N\gt\infty$, $x_{pN}$ satisfies a certain PDE that is easily solvable.
By~(\ref{rwfN}) we have for $n=1,\ldots,N$,
\begin{align*}
&x_N(k+1,n)-x_N(k,n)\nn\\
&=P_r(n-1)x_N(k,n-1)+P_l(n+1)x_N(k,n+1)\nn\\
&\quad-(P_r(n)+P_l(n))x_N(k,n),
\end{align*}
where $x_N(k,n)$ with $n\leq0$ or $n\geq N+1$ are defined to be zero.
Assume $P_l(n)=p_l(nds)$ and $P_r(n)=p_r(nds)$, where $p_l(s)$ and $p_r(s)$ are real-valued functions defined on $\mathcal D$.
Then by
the definition of $x_{pN}$, it follows that for $s\in\mathcal D$ and $t>0$,
\begin{align}
&x_{pN}(t+dt,s)-x_{pN}(t,s)\nn\\
&=p_r(s-ds)x_{pN}(t,s-ds)+p_l(s+ds)x_{pN}(t,s+ds)\nn\\
&\quad-(p_r(s)+p_l(s))x_{pN}(t,s).
\label{equ:xpdiffequ}
\end{align}

To ensure a finite non-degenerate limit, we assume
\[
p_l(s)=b(s)+c_l(s)ds\mbox{ and }p_r(s)=b(s)+c_r(s)ds.
\]
Define $c=c_l-c_r.$ We call $b$ the diffusion coefficient and $c$ the convection coefficient, for a greater $b$ means more rapid diffusion and a greater $c$ means a larger directional bias. Assume that $b\in\mathcal C^2$ and $c\in\mathcal C^1$.
Assume that $x_{pN}$ is twice continuously differentiable in $s$.
Put into (\ref{equ:xpdiffequ}) the Taylor expansions
\begin{align}
x_{pN}(t,s \pm ds)&=\nn x_{pN}(t,s)\pm \frac{\pl x_{pN}}{\pl s}(t,s)ds\\
&\quad+ \frac{\pl^2 x_{pN}}{\pl s^2}(t,s)\frac{ds^2}{2}+o(ds^2),\label{equ:taylor_00}
\end{align}
\begin{align}
b(s \pm ds)= b(s) \pm b_s(s)ds+b_{ss}(s)\frac{ds^2}{2}+o(ds^2),\label{equ:taylor_01}
\end{align}
and
\begin{align}
&c(s \pm ds)= c(s) \pm c_s(s)ds+o(ds),\label{equ:taylor_02}
\end{align}
where a single subscript $s$ represents first derivative and a double subscript $ss$ represents second derivative.
Then we have
\begin{align}
&x_{pN}(t+dt,s)-x_{pN}(t,s)=b(s)\frac{\pl^2 x_{pN}}{\pl s^2}(t,s)ds^2\nn\\
&{}+(2b_s(s)+c(s))\frac{\pl x_{pN}}{\pl s}(t,s)ds^2\nn\\
&{}+(b_{ss}(s)+c_s(s))x_{pN}(t,s)ds^2+o(ds^2).\label{equ:randomwalkode1}
\end{align}
Divide both sides of (\ref{equ:randomwalkode1}) by $dt=ds^2$ and get
\begin{align*}
&\frac{x_{pN}(t+dt,s)-x_{pN}(t,s)}{dt}\nn\\
&=b(s)\frac{\pl^2 x_{pN}}{\pl s^2}(t,s)+(2b_s(s)+c(s))\frac{\pl x_{pN}}{\pl s}(t,s)\nn\\
&\quad+(b_{ss}(s)+c_s(s))x_{pN}(t,s)+\frac{o(ds^2)}{ds^2}. \nn
\end{align*}
As $N\gt \infty$, $ds\gt 0$, and hence $dt=ds^2 \gt 0.$ Assume that $x_{pN}$ is continuously differentiable in $t$. Then by taking the limit as $N\gt \infty$ and rearranging, we get a PDE that $x_{pN}$ satisfies:
\begin{align*}
\dot x_{pN}(t,s)&= \frac{\pl}{\pl s}\left(b(s)\frac{\pl x_{pN}}{\pl s}(t,s)\right)\\
&\quad+\frac{\pl}{\pl s}((b_s(s)+c(s))x_{pN}(t,s)),
\end{align*}
for $t>0$ and $s\in\mathcal D$, with boundary condition $x_{pN}(t,s)=0$.

As $N\gt \infty$, $dt=ds^2\gt 0$, and hence $M=1/dt=1/ds^2\gt\infty$. Then by~(\ref{equ:rwdp}), for $N$ sufficiently large, $X_{pN}$, the continuous-time-space extension of $X_N(k)$, is close to $x_{pN}$, the continuous-time-space extension of $x_N(k)$. Therefore, we can approximate $X_N(k)$ by the solution of the above PDE called the one-dimensional diffusion-convection equation, which can be easily solved~\cite{pdetextbook}. 
Note that our derivation here differs from that of the well-studied Fokker-Planck equation (also known as the Kolmogorov forward equation)~\cite{Gard_SDE}, whereas the latter originates from the study of the probability density of a Wiener process.


This motivational example raises some questions that must be answered by the convergence analysis of the underlying limiting process. First, general networks may exhibit more complex behaviors. For example, $F_N$ might no longer be linear; and SLLN might not apply in many scenarios since node behaviors are not necessarily i.i.d. Specifically, the analysis above does not apply to the network Markov chain in~\cite{chongcontinuum}. To find the conditions under which (\ref{equ:rwdo0}) holds in more general setting, in Section~\ref{subsec:odeconvergence} we apply Kushner's weak convergence theorem in~\cite{kushnerbook} to a more general class of systems modeled by Markov chains. Moreover, we need to show in what sense and under what conditions $X_{pN}$ converges to the solution of the PDE. We analyze such convergence and provide its  sufficient conditions in Section~\ref{subsec:pdeconvergence}.

\section{Continuum Limits of Markov Chains}\label{sec:convergence}
In this section we analyze the convergence of a sequence of Markov
chains to the solution of a PDE in a two-step procedure. We provide
sufficient conditions for this convergence.
\subsection{General Setting}\label{subsec:smmryofmainresult}
Consider $N$ points placed over a Euclidean domain $\mathcal{D}$ representing a spatial
region.
We assume that these points form a \emph{uniform} grid, though our approach can later be generalized to nonuniform cases.
We will refer to these $N$ points in $\mathcal{D}$ as \emph{grid} points and denote the distance between any two neighboring grid points by $ds_N$.


Consider a discrete-time Markov chain
\beq
X_N(k)=[X_N(k,1), \ldots,X_N(k,N)]^T \label{def:MC}
\eeq
with state space $\real^{N}$. Here
$X_N(k,n)$ is the real-valued state of point $n$ at time $k$,
where $n=1,\ldots,N$ is a \emph{spatial} index and $k=0,1,\ldots$ is
a \emph{temporal} index.

Suppose that the evolution of $X_N(k)$ is described by
the stochastic difference equation
\beq
X_N(k+1) = X_N(k) + F_N(X_N(k)/M,U(k)),\label{equ:difference}
\eeq
where $U(k)$ are i.i.d.\ and do not depend on the state $X_N(k)$, $M$ is a
``normalizing'' parameter, and $F_N$ is a given function.
Let
\beq
f_N(x) = EF_N(x,U(k)),\quad x\in\real^N. \label{equ:f_N}
\eeq
Define a deterministic sequence $x_N(k)$ by
\beq
x_N(k+1) = x_N(k) + \frac{1}{M} f_N(x_N(k)), \label{equ:differencenonrand}
\eeq
where $x_N(0)=X_N(0)/M$ a.s.
In the next subsection, we show that under certain conditions, $X_N(k)/M$
and $x_N(k)$ are close in some sense.
\subsection{Convergence to ODE}
\label{subsec:odeconvergence}
Let $X_{oN}(\tld{t})$ be the continuous-time extension of $X_N(k)$ by
piecewise-constant time extensions with interval length $1/M$ and
scaled by $1/M$, i.e., for arbitrary $\tld{t}\in\real$,
\beq
X_{oN}(\tld{t}) = X_N(\floor{M\tld{t}})/M. \label{equ:X_{oN}def}
\eeq
It follows that for each $k$, $X_{oN}(k/M) = X_N(k)/M$.
Similarly we define $x_{oN}(\tld{t})$, the continuous-time extension of $x_N(k)$ by
\beq
x_{oN}(\tld{t}) = x_N(\floor{M\tld{t}}). \label{equ:x_{oN}def}
\eeq

For fixed $\tld {T}_N>0$, let $D^N[0,\tld {T}_N]$ be the space of
$\real^N$-valued C\`{a}dl\`{a}g functions on $[0,\tld {T}_N]$, i.e.,
functions that are right-continuous at each $t\in[0,\tld {T}_N)$ and have left-hand limits at each $t \in (0,\tld {T}_N]$. As defined in (\ref{equ:X_{oN}def}) and (\ref{equ:x_{oN}def}) respectively, both $X_{oN}(\tld{t})$ and $x_{oN}(\tld{t})$ with $\tld{t} \in [0,\tld {T}_N]$ are in $D^N[0,\tld {T}_N].$
Since both  $X_{oN}(\tld{t})$ and $x_{oN}(\tld{t})$ depend on $M$, each one of them forms a sequence of functions in $D^N[0,\tld {T}_N]$ indexed by $M=1,2,\ldots$.

Define the $\infty$-norm $\|\cdot\|_\infty^{(o)}$ on $D^N[0,\tld T_N]$, i.e., for $x\in D^N[0,\tld T_N]$,
\[
\|x\|_\infty^{(o)}=\max_{n=1,\ldots,N}\sup_{t\in [0,\tld {T}_N]}|x^n(t)|,
\]
where $x^n$ is the $n$th components of $x$.
A sequence of functions $x_M \in D^N[0,\tld T_N]$ is said to converge uniformly to a function $x\in D^N[0,\tld T_N]$ if as $M\gt\infty$, $\|x_M-x\|_\infty^{(o)}\to0$. In this paper, we use the notation ``$\Rightarrow$'' for weak convergence and ``$\xrightarrow{P}$'' for convergence in probability.

Let $f_N$ be defined as in (\ref{equ:f_N}). Now we present a lemma
stating that under some conditions, as $M \gt \infty$, $X_{oN}$ converges uniformly to a limiting function $y$, the solution of the ODE $\dot{y}=f_N(y)$, on $[0,\tld {T}_N]$, and $X_{oN}$ converges uniformly to the same solution on $[0,\tld {T}_N]$.
\blemma \label{lmm:ourodeconv}
Assume:
\ben
\item[(1a)] There exists an identically distributed sequence $\{\lambda(k)\}$ of integrable random variables such that for each $k$ and $x$, $|F_N(x,U(k))|\leq \lambda(k)$ a.s.;
\item[(1b)] the function $F_N(x,U(k))$ is continuous in $x$ a.s.; and
\item[(1c)] the ODE $\dot{y}=f_N(y)$ has a unique solution on $[0,\tld {T}_N]$ for any initial condition $y(0)$.
\een
Suppose that as $M\to\infty$,
\[
X_{oN}(0)\xrightarrow{P} y(0) \mbox{ and } x_{oN}(0)\to y(0).
\]
Then, as $M\to\infty$,
\[
\|X_{oN} -y\|_\infty^{(o)}\xrightarrow{P}0 \mbox{ and } \|x_{oN} -y\|_\infty^{(o)}\to0
\]
on $[0,\tld {T}_N]$, where $y$ is the unique solution of $\dot{y}=f_N(y)$ with initial condition $y(0)$.
\elemma

To prove Lemma~\ref{lmm:ourodeconv}, we first present a lemma on weak convergence due to Kushner~\cite{kushnerbook}.
\blemma \label{lmm:kushnerweakconv}
Assume:
\ben
\item[(2a)]
The set 
\[
\{|F_N(x, U(k))|: k\geq 0\}
\]
is uniformly integrable;
\item[(2b)]  for each $k$ and each bounded random variable $X$,
\[
\lim_{\delta\gt 0} E\sup_{|Y|\leq \delta}|F_N(X, U(k))-F_N(X+Y,U(k))| = 0;
\]
and
\item[(2c)] there is a function $\hat{f}_N(\cdot)$ \textup{[}continuous by~(b)\textup{]} such that as $n \gt \infty$,
\[
\frac{1}{n}\sum^{n}_{k=0}{F_N(x, U(k))\xrightarrow{P} \hat{f}_N(x)}.
\]
\een
Suppose that $\dot{y}=\hat{f}_N(y)$ has a unique solution on $[0,
\tld T_N]$ for each initial condition, and that $X_{oN}(0) \Rightarrow y(0)$. Then as $M\to\infty$,
\[
\|X_{oN}-y\|_\infty^{(o)}\Rightarrow 0 \mbox{ on } [0,\tld T_N].
\]
\elemma
We note that in Kushner's work, the convergence of $X_{oN}$ to $y$ is stated in terms of Skorokhod norm~\cite{kushnerbook}, but it is equivalent to the $\infty$-norm in our case where the functions are defined on finite time intervals~\cite{kushner_yin}.

We now prove Lemma~\ref{lmm:ourodeconv} by showing that the assumptions {(2a)--(2c)} in Lemma~\ref{lmm:kushnerweakconv} hold under the assumptions {(1a)--(1c)} in Lemma~\ref{lmm:ourodeconv}.
\bproofof{Lemma~\ref{lmm:ourodeconv}}
\ben
\item Since $\lambda(k)$ is integrable, as $a\gt\infty$,
\[
E |\lambda(k)|1_{\{|\lambda(k)|>a\}} \gt 0,
\]
where $1_A$ is the indicator function of set $A$.
By Assumption {(1a)}, for each $k$ and $x$,
\[
F_N(x,U(k))\leq \lambda(k) \mbox{ a.s.}
\]
Therefore for each $x$ and $a>0$,
\begin{align}
\lefteqn{E |F_N(x,U(k))|1_{\{|F_N(x,U(k))|>a\}}}\nn\\
&\leq E |\lambda(k)|1_{\{|F_N(x,U(k))|>a\}}\nn\\
&\leq E |\lambda(k)|1_{\{|\lambda(k)|>a\}}.\nn
\end{align}
Hence as $a\gt\infty$,
\[
\sup_{k\geq 0} E|F_N(x,U(k))|1_{\{|F_N(x,U(k))|>a\}} \gt 0,
\]
i.e., the family $\{|F_N(x,U(k))|: k\geq 0\}$ is uniformly integrable and Assumption {(2a)} holds.

\item
By Assumption {(1b)}, $F_N(x,U(k))$ is continuous in $x$ a.s. Then for each bounded $X$ and each $k$,
\[
\lim_{\delta \gt 0}\sup_{|Y|\leq \delta}|F_N(X,U(k))-F_N(X+Y,U(k))|=0 \mbox{ a.s.}
\]
By Assumption {(1a)}, for each $x$ and each $k$, there exists an integrable random variable $\lambda(k)$ such that $|F_N(x,U(k))|\leq \lambda(k)$ a.s.
It follows that for each bounded $X$, each $k$, and each $Y$ such that $|Y|\leq\delta$,
\begin{align}
&|F_N(X,U(k))-F_N(X+Y,U(k))|\nn\\
&\leq |F_N(X,U(k))|+|F_N(X+Y,U(k))|\leq 2\lambda(k)\nn.
\end{align}
Hence for each $\delta$,
\[
\left|\sup_{|Y|\leq \delta} |F_N(X,U(k))-F_N(X+Y,U(k))|\right|\leq 2\lambda(k),
\]
an integrable random variable.
By the dominant convergence theorem,
\begin{align}
&\lim_{\delta \gt 0}E\sup_{|Y|\leq \delta}|F_N(X,U(k))-F_N(X+Y,U(k))| \nn\\
&=E \lim_{\delta \gt 0}\sup_{|Y|\leq\delta}|F_N(X,U(k))-F_N(X+Y,U(k))|\nn\\
&=0.\nn
\end{align}
Hence Assumption {(2b)} holds.
\item
Since $U(k)$ are i.i.d., by the weak law of large numbers and the
definition of $f_N$ in (\ref{equ:f_N}),
as $n \gt \infty$,
\[
\frac{1}{n}\sum^{n}_{k=0}{F_N(x, U(k))\xrightarrow{P}f_N(x)}.
\]
Hence Assumption {(2c)} holds.
\een
Then, by Lemma~\ref{lmm:kushnerweakconv}, as $M\to\infty$,  $\|X_{oN}- y\|_\infty^{(o)}\Rightarrow 0$ on $[0,\tld T_N]$. For each sequence of random processes $\{X_n\}$, if $A$ is a constant, $X_n \Rightarrow A$ if and only if $X_n \xrightarrow{P} A$. Therefore, as $M\to\infty$, \mbox{$\|X_{oN}-y\|_\infty^{(o)}$}$\xrightarrow{P}0$ on $[0,\tld T_N]$.
The same argument implies the deterministic convergence of $x_{oN}$: as $M\gt\infty$, $\|x_{oN} - y\|_\infty^{(o)}\to0$ on $[0,\tld T_N]$.
\eproofof

Based on Lemma~\ref{lmm:ourodeconv}, we get the following lemma,
which states that $X_{oN}$ and $x_{oN}$ are close with high probability when $M$ is large.
\blemma \label{lmm:smalldo}
Let the assumptions in Lemma~\ref{lmm:ourodeconv} hold.
Then for any sequence $\{\zeta_N\}$, for each $N$, and for $M$ sufficiently large, we have
\beq
P\{\|X_{oN} - x_{oN}\|_\infty^{(o)} > \zeta_N \} \leq 1/N^2\mbox{ on } [0,\tld {T}_N].\nn
\eeq
\elemma
\bproof
By Lemma~\ref{lmm:ourodeconv}, for each $N$, as $M\to\infty$,
\[
\|X_{oN} - y\|_\infty^{(o)}\xrightarrow{P}0 \mbox{ and } \|x_{oN} - y\|_\infty^{(o)}\to0 \mbox{ on } [0,\tld {T}_N].
\]
By the triangle inequality
\[\|X_{oN} - x_{oN}\|_\infty^{(o)} \leq \|X_{oN} - y\|_\infty^{(o)} + \|x_{oN} - y\|_\infty^{(o)},\]
it follows that as $M\to\infty$, $\|X_{oN} - x_{oN}\|_\infty^{(o)} \xrightarrow{P} 0$ on $[0,\tld {T}_N]$.
This finishes the proof.
\eproof

Since $X_{oN}$ and $x_{oN}$ are the piecewise continuous-time extensions of $X_{N}$ and $x_{N}$ by constant interpolation, respectively, we have the following corollary.
\bcor\label{cor:ODE}
Fix $\tld {T}_N$ and let $\tld K_N=\floor{\tld T_NM}$.
Let the assumptions in Lemma~\ref{lmm:ourodeconv} hold.
Then for any sequence $\{\zeta_N\}$, for each $N$, and for $M$ sufficiently large, we have
\beq
P\left\{\max_{{k=0,\ldots,\tld K_N}\atop{n=1,\ldots,N}}\left|\frac{X_N(k,n)}{M}-x_N(k,n)\right| > \zeta_N \right\} \leq \frac{1}{N^2}.\nn
\eeq
\ecor
We use Lemma~\ref{lmm:smalldo} and Corollary~\ref{cor:ODE} in the next subsection.
\subsection{Convergence to PDE}
\label{subsec:pdeconvergence}
In the last subsection, we stated conditions under which the continuous-time extensions of
$X_N(k)$ and $x_N(k)$ are close asymptotically (as $M\to\infty$)
with high probability.
In this subsection, we further let $N\to\infty$ and state conditions under which  $x_{N}(k)$  is close
asymptotically to the solution of a PDE. This leads to the convergence of $X_N(k)/M$ to the PDE solution as $M\to\infty$ and $N\to\infty$.

Assume that the domain $\mathcal{D}$ introduced in
Section~\ref{subsec:smmryofmainresult} is compact and convex,
and let $w:\mathcal{D}\to\real$ be in $\mathcal C^2$.
Given a fixed $N$, let $V_N$ be the set of the $N$ grid points in
$\mathcal{D}$.  Let $y_N$ be the vector in $\real^N$ composed of the values of $w$
at the grid points $v_N(n)\in V_N$, $n=1,\ldots,N$, i.e.,
$y_N=[w(v_N(1)),\ldots,w(v_N(N))]^T.$

Given $s\in \mathcal{D}$, let $\{s_N\}\subset \mathcal{D}$ be a sequence of grid points in $\mathcal{D}$ such that as $N \gt \infty$, $s_N\gt s$, where for each $N$, $s_N$ is a grid point in $V_N$.
Let $f_N(y_N, s_N)$ be the component of the vector $f_N(y_N)$ corresponding to the location $s_N$. For example, for $N=5$, if $s_5=v_5(4)$ in $V_5$, then $f_5(y_5, s_5)$ is the 4th component of the vector $f_5(y_5)$.

Assume that there exist sequences $\{\delta_N\}$, $\{\beta_N\}$, $\{\gamma_N\}$, and $\{\rho_N\}$, functions $f$ and $h$, and $0<c<\infty$, such that as $N\gt \infty, \delta_N\gt 0,\delta_N/\beta_N\gt 0, \gamma_N\to0, \rho_N\to0$, and:
\bit
\item for any $s_N$ such that $s_N\to s$, where $s$ is in the interior of $\mathcal{D}$, there exists a sequence of functions $\phi_N:\mathcal D\to\real$ such that
\begin{align}
f_N(y_N,s_N)/\delta_N &= f(s_N, w(s_N) ,\nabla w(s_N), \nabla^2 w(s_N))\nn\\
&\quad+\phi_N(s_N),\label{equ:w_diff1}
\end{align}
and for $N$ sufficiently large,
\beq
|\phi_N(s_N)|\leq c\gamma_N; \label{def:gamma_N}
\eeq
and
\item for any $s_N$ such that $s_N\to s$, where $s$ is on the boundary of $\mathcal{D}$, there exists a sequence of functions $\varphi_N:\mathcal D\to\real$ such that
\begin{align}
f_N(y_N,s_N)/\beta_N &= h(s_N, w(s_N) ,\nabla w(s_N), \nabla^2 w(s_N))\nn\\
&\quad+\varphi_N(s_N),\label{equ:w_diff2}
\end{align}
and for $N$ sufficiently large, $|\varphi_N(s_N)|\leq c\rho_N$.
\eit
Here, $\nabla^i w$ represents all the $i$th order
derivatives of $w$, where $i=1,2$.

These assumptions are technical conditions on the asymptotic
behavior of the sequence of functions $f_N$. The basic idea is that
$f_N(y_N,s_N)$ is asymptotically close to some function of terms
that look like the right-hand side of a time-dependent PDE.
Typically, checking these conditions amounts to simply an algebraic
exercise. A concrete example of this is given in the next section.

The basic idea underlying the analysis in the remainder of this
subsection is this.
Recall that $x_N(k)$ is defined by~(\ref{equ:differencenonrand}).
Suppose we associate the discrete time $k$ with points on the real
line spaced apart by a distance proportional to $\delta_N$. Then,
the above technical assumption implies that $x_N(k)$ is, in some
sense, close to the solution of a PDE of the form
$\dot{z}=f(s, z ,\nabla z, \nabla^2 z)$ with boundary
condition $h(s, z ,\nabla z, \nabla^2 z)=0$.
Because the Markov chain $X_N(k)/M$ is close to $x_N(k)$, as established in the
last subsection, it is also close to the solution of the PDE.
The remainder of this subsection is devoted to developing this argument rigorously.

Fix $T>0$. Assume that there exists a function $z:[0,T]\times \mathcal{D} \to\real$ that solves the PDE
\beq\dot{z}(t,s)=  f(s, z(t,s) ,\nabla z(t,s), \nabla^2 z(t,s)),\label{equ:z_pde1}
\eeq
with boundary condition
\beq\ h(s, z(t,s) ,\nabla z(t,s) \nabla^2 z(t,s))=0\nn
\eeq
and initial condition $z(0,s)=z_0(s)$.
Here, $\nabla^i z(t,s)$ represents all the $i$th order partial derivatives of $z(t,s)$ with respect to $s$, where $i=1,2$.


Define
\beq
dt_N=\delta_N/M. \label{equ:def_dt}
\eeq
Define
\[
K_N=\floor{T/dt_N}\mbox{ and } t_N(k)=kdt_N.
\]
Define
\[
z_N(k,n)=z(t_N(k),v_N(n))
\]
and let $z_N(k)=[z_N(k,1),\ldots,z_N(k,N)]^T\in\real^N$.

Denote the $\infty$-norm on $\real^N$ by $\|\cdot\|^{(N)}_\infty$.
That is, for $x\in \real^N$, with the $n$th element being $x(n)$,
\[
\|x\|^{(N)}_\infty=\max_{1\leq n\leq N}|x(n)|.
\]
Denote the $\infty$-norm on $\real^{N\times K_N}$ also by $\|\cdot\|^{(N)}_\infty$. That is, for $x=[x(1),\ldots,x(K_N)]\in\real^{N\times K_N}$, where for $k=1,\ldots,K_N$, $x(k)=[x(k,1),\ldots,x(k,N)]^T\in\real^N$, we have
\begin{align*}
\|x\|_\infty^{(N)}=\max_{{k=1,\ldots,K_N}\atop{n=1,\ldots,N}}|x(k,n)|.
\end{align*}

Now we present a lemma on the relationship between the $z_N(k)$ and $f_N$.
\blemma \label{lmm:difference_z}
Assume that $z$ is continuously differentiable in $t$.
Then for each $N$, there exists $u_N(k)\in\real^N$ such that for $k=0,\ldots,K_N-1$,
\beq
z_N(k+1)-z_N(k)=\frac{1}{M}f_N(z_N(k))+dt_Nu_N(k),\label{equ:z_diff}
\eeq
and
\beq
\|u_N\|^{(N)}_\infty=O(\max\{\gamma_N,dt_N\}),\label{equ:u_to0}
\eeq
where $u_N=[u_N(0),\ldots,u_N(K_N-1)]\in\real^{N\times K_N}.$
\elemma
\bproof
Since $z$ is continuously differentiable in $t$, there exists $0<c_1<\infty$ such that for each $N$, for $k=0,\ldots,K_N-1$ and $n=1,\ldots,N$, there exists a function $r_N:[0,T]\times\mathcal D\to\real$ such that
\begin{align}
&\frac{z_N(k+1,n)-z_N(k,n)}{dt_N}\nn\\
&=\frac{z(t_N(k),v_N(n))-z(t_N(k),v_N(n))}{dt_N}\nn\\
&=\dot z(t_N(k),v_N(n))+r_N(t_N(k),v_N(n)),\label{equ:z_N_in1}
\end{align}
and for $N$ sufficiently large, $|r_N(t_N(k),v_N(n))|<c_1dt_N$.

By~(\ref{equ:w_diff1}) and (\ref{equ:z_pde1}), there exists
$0<c_2<\infty$, such that for each $N$, for $k=0,\ldots,K_N-1$ and $n=1,\ldots,N$, there exists a function $\phi_N:[0,T]\times\mathcal D\to\real$ such that
\begin{align}
&\dot z(t_N(k),v_N(n))\nn\\
&=f(v_N(n),z_N(k,n),\nabla z_N(k,n),\nabla^2 z_N(k,n))\nn\\
&=f_N(z_N(k),v_N(n))/\delta_N+\phi_N(t_N(k),v_N(n)),\label{equ:z_N_in2}
\end{align}
and  for $N$ sufficiently large, $|\phi_N(t_N(k),v_N(n))|<c_2\gamma_N$, where $\{\gamma_N\}$ is as defined in~(\ref{def:gamma_N}).

For each $N$, for $k=0,\ldots,K_N-1$ and $n=1,\ldots,N$, let
\beq
u_N(k,n)=\phi_N(t_N(k),v_N(n))+r_N(t_N(k),v_N(n)),\nn
\eeq
and $u_N(k)=[u_N(k,1),\ldots,u_N(k,N)]^T\in\real^N$.
Then there exists $0<c<\infty$ such that for each $N$,
\[
\|u_N\|^{(N)}_\infty<c\max\{\gamma_N,dt_N\}.
\]
Hence (\ref{equ:u_to0}) follows.

By~(\ref{equ:z_N_in1}) and (\ref{equ:z_N_in2}), for each $N$, for $k=0,\ldots,K_N-1$ and $n=1,\ldots,N$,
\beq
\frac{z_N(k+1)-z_N(k)}{dt_N}=\frac{f_N(z_N(k))}{\delta_N}+u_N(k).\nn
\eeq
By this and~(\ref{equ:def_dt}), we have (\ref{equ:z_diff}).
\eproof

In the following we show that under some conditions, $x_N(k)$ and $z_N(k)$ are asymptotically close for large $N$.

For each $N$, for $k=0,\ldots,K_N$ and $n=1,\ldots,N$, define
\beq
\ep_N(k,n)=z_N(k,n)-x_N(k,n),\label{equ:def_ep}
\eeq
and let $\ep_N(k)=[\ep_N(k,1),\ldots,\ep_N(k,N)]^T\in\real^N$.

By~(\ref{equ:differencenonrand}), (\ref{equ:z_diff}), and (\ref{equ:def_ep}), we have that for each $N$, for $k=0,\ldots,K_N$, there exists $u_N(k)$ as defined in Lemma~\ref{lmm:difference_z} such that
\begin{align}
\ep_N(k+1)&=\ep_N(k)+\frac{1}{M}(f_N(z_N(k))-f_N(x_N(k)))\nn\\
&\quad+dt_Nu_N(k).\label{equ:ep}
\end{align}

Suppose that for each $N$, $f_N\in\mathcal C^1$. Let $Df_N(x)$ be the derivative matrix of the function $f_N$ at $x$. Then we have that for each $N$, for $k=1,\ldots,K_N$ and $n=1,\ldots,N$, there exists a function $\tld f_N:\real^N\to\real^N$ such that
\beq
f_N(z_N(k))-f_N(x_N(k))=Df_N(z_N(k))\ep_N(k)+ \tld f_{N}(\ep_N(k))\nn
\eeq
and
\beq
\tld f_N(0)=0.\label{equ:tld_f_N_0_is_0}
\eeq
Then we have from~(\ref{equ:ep})
\begin{align}
\ep_N(k+1)&=\ep_N(k)
+\frac{1}{M}(Df_N(z_N(k))\ep_N(k)\nn\\
&\quad+\tld f_{N}(\ep_N(k)))+dt_Nu_N(k).\label{equ:ep1}
\end{align}
Further suppose that for each $N$,
\beq
\|\ep_N(0)\|_\infty^{(N)}=0.\label{equ:ep0_0}
\eeq
Define $\ep_N=[\ep_N(1),\ldots,\ep_N(K_N)]\in \real^{N\times K_N}$.
Then by~ (\ref{equ:tld_f_N_0_is_0}), (\ref{equ:ep1}), and (\ref{equ:ep0_0}), for each $N$, there exists a function $H_N:\real^{N\times K_N}\to\real^{N\times K_N}$ such that
\begin{align}
\ep_N=H_N(u_N).\label{equ:error_sys_H}
\end{align}
It follows that $H_N(0)=0$ and $H_N\in\mathcal C^1$.

For each $N$, define
\begin{align}
\mu_N=\lim_{\alpha\to0}\sup_{\|u \|_\infty^{(N)}\leq\alpha}\frac{\|H_N(u)\|_\infty^{(N)}}{\|u \|_\infty^{(N)}}.\nn
\end{align}
\begin{lemma}\label{lmm:h_N}
Assume that
\bit
\item $z$ is continuously differentiable in $t$;
\item for each $N$, $f_N\in\mathcal C^1$;
\item for each $N$, (\ref{equ:ep0_0}) holds; and
\item the sequence $\{\mu_N\}$ is bounded.
\eit
Then
\[\|\ep_N \|_\infty^{(N)}=O(\|u_N \|_\infty^{(N)}).\]
\end{lemma}
\bproof
By definition, for each $N$, there exists $\delta>0$ such that for $\alpha<\delta$,
\begin{align*}
\sup_{\|u \|_\infty^{(N)}\leq\alpha}\frac{\|H_N(u)\|_\infty^{(N)}}{\|u \|_\infty^{(N)}}<\mu_N+1.
\end{align*}
By~(\ref{equ:u_to0}), as $N\to\infty$, $\|u \|_\infty^{(N)}\to0$. Then there exists $N_0$ and $\alpha_1$ such that for $N>N_0$, $\|u \|_\infty^{(N)}\leq\alpha_1<\delta$.
Hence, for $N>N_0$,
\begin{align*}
\frac{\|H_N(u)\|_\infty^{(N)}}{\|u \|_\infty^{(N)}}\leq
\sup_{\|u \|_\infty^{(N)}\leq\alpha_1}\frac{\|H_N(u)\|_\infty^{(N)}}{\|u \|_\infty^{(N)}}<\mu_N+1.
\end{align*}
Therefore, there exists $0<c<\infty$ such that for $N>N_0$,
\begin{align*}
\|\ep_N\|_\infty^{(N)}&=\|H_N(u_N)\|_\infty^{(N)}<(\mu_N+1)\|u_N\|_\infty^{(N)}\\
&<(c+1)\|u_N \|_\infty^{(N)}.
\end{align*}
This finishes the proof.
\eproof
Lemma~\ref{lmm:h_N} states that as $N\to\infty$,  $\|\ep_N\|_\infty^{(N)}\to0$, and at least with the same rate as $\|u_N\|_\infty^{(N)}$.

Let $X_N=[X_N(1)/M,\ldots,X_N(K_N)/M]$,  $x_N=[x_N(1),\ldots,x_N(K_N)]$, and $z_N=[z_N(1),\ldots,z_N(K_N)]$, all in $\in\real^{N\times K_N}$. Now we present the main convergence theorem of this paper, which states that the value of the normalized Markov chain at time $k$ and node $n$, is close to that of $z$ at the corresponding point $(t_N(k),v_N(n))\in[0,T]\times\mathcal D$ for large $M$ and $N$.
\bthm \label{thm:main_thm}
Suppose that the assumptions in Lemma~\ref{lmm:ourodeconv} and Lemma~\ref{lmm:h_N} hold.
Then
\begin{align*}
&\|X_N-z_N\|_\infty^{(N)}=O(\max\{\gamma_N,dt_N\}) \mbox{ a.s.}
\end{align*}
\ethm
\bproof
By~(\ref{equ:u_to0}) and Lemma~\ref{lmm:h_N}, there exists $0<c_0<\infty$ such that for $N$ sufficiently large,
\beq
\|\ep_N \|_\infty^{(N)}<c_0\max\{\gamma_N,dt_N\}.\label{inequ:ep}
\eeq

Let $\tld {T}_N$ in Corollary~\ref{cor:ODE} be $T/\delta_N$. Then $\tld K_N:=\floor{\tld {T}_NM}=\floor{T/dt_N}:=K_N$. Hence by Corollary~\ref{cor:ODE}, for any sequence $\{\zeta_N\}$, for each $N$,  we can take $M$ sufficiently large such that
\beq
\sum_{N=1}^{\infty}P\{\|X_N-x_N\|_\infty^{(N)} > \zeta_N \}\leq\sum_{N=1}^{\infty}1/N^2<\infty.\nn
\eeq
By the first Borel-Cantelli Lemma~\cite{statbook},
\beq
P\left\{\lim\sup_{N \gt \infty}\{\|X_N-x_N\|_\infty^{(N)} > \zeta_N\}\right\}=0,\nn
\eeq
which implies that, a.s., for $N$ sufficiently large,
\[
\|X_N-x_N\|_\infty^{(N)} < \zeta_N.
\]
Take $\zeta_N$ such that for $N$ sufficiently large, 
\[
\zeta_N<\max\{\gamma_N,dt_N\}.
\]
Then by the triangle inequality
\begin{align*}
\|X_N-z_N\|_\infty^{(N)} &\leq \|X_N-x_N\|_\infty^{(N)}+\|x_N-z_N\|_\infty^{(N)}\\
&=\|X_N-x_N\|_\infty^{(N)}+\|\ep_N \|_\infty^{(N)},
\end{align*}
a.s., there exists $0<c<\infty$ such that for $N$ sufficiently large,
\beq
\|X_N-z_N\|_\infty^{(N)}\leq c\max\{\gamma_N,dt_N\}.\nn
\eeq
This finishes the proof.
\eproof
This theorem states that as $M \gt \infty$ and $N \gt \infty$, $X_N$ converges uniformly to $z_N$ a.s., and at least with the same rate as $\max\{\gamma_N,dt_N\}$.
\subsection{Convergence of Continuous-time-space Extension}
\begin{figure}
\centering
\includegraphics[width=60mm]{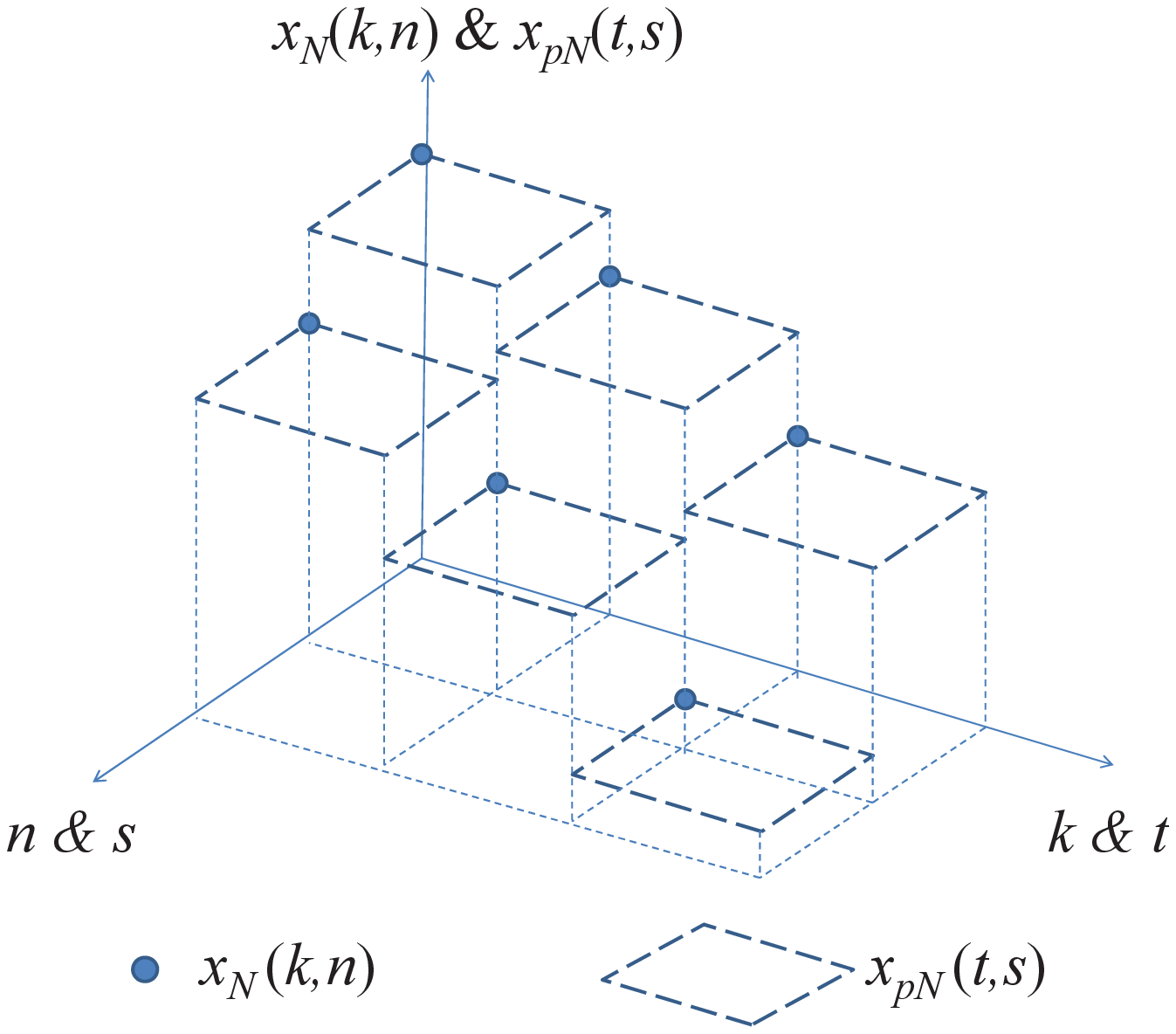}\\
\caption{An illustration of $x_N$ and $x_{pN}$ in a one-dimensional case.}
\label{fig:xN_xpN}
\end{figure}

In the following we study the convergence of the continuous-time-space extension of the Markov chain $X_N(k)$ to the PDE solution.
Set $\tld {T}_N=T/\delta_N$. For each $N$, we can construct $X_{oN}(\tld {t})$ and $x_{oN}(\tld {t})$ with time interval of length $1/M$, with
$\tld{t}\in [0,\tld {T}_N]$.
Respectively, let $X_{pN}(t)$ and $x_{pN}(t)$, where $t\in[0,T]$, be the continuous-space
extension of $X_{oN}(\tld {t})$ and $x_{oN}(\tilde{t})$ (with
$\tld{t}\in [0,\tld {T}_N]$)
by piecewise-constant space extensions on $\mathcal{D}$ and
with time scaled by $\delta_N$ so that the time-interval length is
$\delta_N/M:= dt_N$. By \emph{piecewise-constant space extension} of
$X_{oN}$, we mean that we construct a piecewise-constant function on
$\mathcal{D}$ such that the value of this function at each point in $\mathcal{D}$ is the
value of the component of the vector $X_{oN}$ corresponding to the grid
point that is ``closest to the left'' (taken one component at a
time). Then for each $t$, $X_{pN}(t)$ and $x_{pN}(t)$ are real-valued functions defined on $\mathcal{D}$. Fig.~\ref{fig:xN_xpN} is an illustration of $x_N$ and $x_{pN}$ in a one-dimensional case.
For fixed $T$, both $X_{pN}(t)$ and $x_{pN}(t)$ with $t\in[0,T]$ are in the space $D^{\mathcal{D}}[0,T]$ of functions of  $[0,T]\times\mathcal D \to \real$ and are C\`{a}dl\`{a}g with the time component.
Define the $\infty$-norm $\|\cdot\|_\infty^{(p)}$ on $D^{\mathcal{D}}[0,T]$, i.e., for $x\in D^{\mathcal{D}}[0,T]$,
\[
\|x\|_\infty^{(p)}=\sup_{t\in [0,T],\atop s\in\mathcal D}|x(t,s)|.
\]

First we show that $x_{pN}$ and $z$ are asymptotically close for large $N$. 
\blemma \label{lmm:PDEconv}
Suppose that the assumptions in Lemma~\ref{lmm:h_N} hold. Then
\beq
\|x_{pN}-z\|_\infty^{(p)}=O(\max\{\gamma_N,dt_N,ds_N\}).\nn
\eeq
\elemma
\bproof
For each $N$, for $k=0,\ldots,K_N$ and $n=1,\ldots,N$, by the definition of $x_{pN}$, we have that  $x_{pN}(t_N(k),v_N(n))=x_N(k,n)$. Let $\Omega_N(k,n)$ be the
subset of $[0,T]\times \mathcal{D}$ containing $(t_N(k),v_N(n))$ where $x_{pN}$ is piecewise constant, i.e., $(t_N(k),v_N(n))\in\Omega_N(k,n)$ and for all $(t,s)\in\Omega_N(k,n)$, $x_{pN}(t,s)=x_{pN}(t_N(k),v_N(n))$. (For example, for $\mathcal{D}\subset\real$, $\Omega_N(k,n)=[t_N(k),t_N(k+1)]\times[v_N(n),v_N(n+1)]$.)
Then for each $N$,
\begin{align}
&\|x_{pN}-z\|_\infty^{(p)}\leq \|\ep_N \|_\infty^{(N)}\nn\\
&\quad+\max_{k=0,\ldots,K_N\atop n=1,\ldots,N} \sup_{(t,s)\in\Omega_N(k,n)}
|z(t_N(k),v_N(n))-z(t,s)|.\nn
\end{align}

Since $z(t,s)$ is continuously differentiable in $t$ on a compact domain, it is Lipschitz continuous in $t$.
Similarly, it is Lipschitz continuous in $s$.
Hence there exist $0<c_1,c_2\leq\infty$ such that for each $N$,
\begin{align}
&\max_{k=0,\ldots,K_N,\atop n=1,\ldots,N} \sup_{(t,s)\in\Omega_N(k,n)}
|z(t_N(k),v_N(n))-z(t,s)|\nn\\
&\leq c_1\max_{k=0,\ldots,K_N,\atop n=1,\ldots,N} \sup_{(t,s)\in\Omega_N(k,n)}
\|(t_N(k),v_N(n))-(t,s)\|\nn\\
&\leq c_2\max\{ds_N,dt_N\},\nn
\end{align}
where $\|\cdot\|$ is some norm on $[0,T]\times \mathcal{D}$.
Hence, by this and~(\ref{inequ:ep}), there exists $0<c<\infty$ such that for $N$ sufficiently large,
\beq
\|x_{pN}-z\|_\infty^{(p)}\leq c\max\{\gamma_N,dt_N,ds_N\}.\nn
\eeq
This finishes the proof.
\eproof

Now we present a convergence theorem for the continuous functions.
\bthm \label{thm:oldmainthm}
Suppose that the assumptions in Lemma~\ref{lmm:ourodeconv} and Lemma~\ref{lmm:h_N} hold.
Then
\beq
\|X_{pN} - z \|_\infty^{(p)}=O(\max\{\gamma_N,dt_N,ds_N\})\mbox{ a.s.\ on } [0,T]\times \mathcal{D}.\nn
\eeq
\ethm
\bproof
By Lemma~\ref{lmm:smalldo}
, for any sequence $\{\zeta_N\}$, for each $N$,  we can take $M$ sufficiently large such that
\[
\sum_{N=1}^{\infty}P\{\|X_{oN}- x_{oN}\|_\infty^{(o)} > \zeta_N \}\leq\sum_{N=1}^{\infty}1/N^2<\infty.
\]
By the first Borel-Cantelli Lemma~\cite{statbook},
\[
P\left\{\lim\sup_{N \gt \infty}\{\|X_{oN}- x_{oN}\|_\infty^{(o)} > \zeta_N\ \}\right\}=0,
\]
which implies that, a.s., for $N$ sufficiently large,
\beq
\|X_{oN}- x_{oN}\|_\infty^{(o)} < \zeta_N \mbox{ on }[0,\tld T_N].\nn
\eeq
Since $X_{pN}$ and $x_{pN}$ are the piecewise continuous-space extensions of $X_{oN}$ and $x_{oN}$ by constant interpolation, respectively, it follows that for any sequence $\{\zeta_N\}$, we can take $M$ sufficiently large such that, a.s.,
for $N$ sufficiently large,
\[
\|X_{pN}- x_{pN}\|_\infty^{(p)} < \zeta_N  \mbox{ on }[0,T]\times \mathcal{D}.
\]
Take $\zeta_N$ such that for $N$ sufficiently large, 
\[
\zeta_N<\max\{\gamma_N,dt_N,ds_N\}.
\]
Then by the triangle inequality
\[
\|X_{pN}-z\|_\infty^{(p)} \leq \|X_{pN}-x_{pN}\|_\infty^{(p)} + \|x_{pN}-z\|_\infty^{(p)}
\]
and Lemma~\ref{lmm:PDEconv}, a.s.,
there exists $0<c<\infty$ such that for $N$ sufficiently large,
\beq
\|X_{pN} - z \|_\infty^{(p)}\leq c\max\{\gamma_N,dt_N,ds_N\}\mbox{ on }[0,T]\times \mathcal{D}.\nn
\eeq
This finishes the proof.
\eproof
This theorem states that as $M \gt \infty$ and $N \gt \infty$, the
continuous-time-space extension $X_{pN}$ of the Markov chain $X_N(k)$, converges uniformly to $z$, the solution of the PDE a.s., and at least with the same rate as $\max\{\gamma_N,dt_N,ds_N\}$.

The solution of the PDE can be
found quickly by mathematical tools readily available and then be used
to approximate the Markov chain $X_N(k)$. We give an example of this in the next section.
\section{Application to the Modeling of Large Networks}
\label{sec:networkexample}
In this section we present an example of the application of our
approach to network modeling. We show how the Markov chain
representing the queue lengths of the nodes in
the network can be approximated by the solution of a PDE using
the results of the preceding section.

\subsection{Network Model}
\begin{figure}
\centering
\includegraphics[width=60mm]{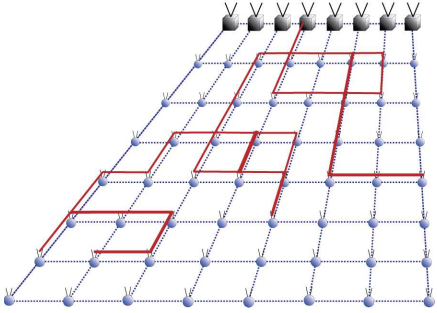}
\caption{An illustration of a wireless sensor network over a two-dimensional domain. Destination nodes are located at the far edge. We show the possible path of a message originating from a node located in the left-front region.}
\label{fig:sensornetwork}
\end{figure}

\begin{figure}
\centering
\includegraphics[width=50mm]{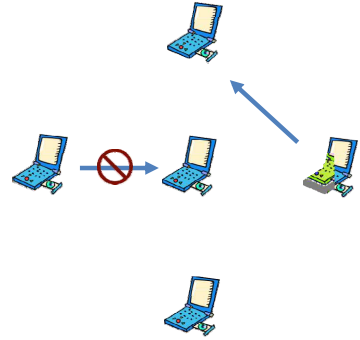}
\caption{An illustration of the collision protocol: reception at a
node fails when more than one of its neighbors transmit (regardless of the intended receiver).}
\label{fig:macrule}
\end{figure}

We consider a network of wireless sensor nodes uniformly placed over a domain. In a random fashion, the sensor nodes generate data messages that need to be communicated to the destination nodes located on the boundary of the domain, which represent specialized devices that collect the sensor data. The sensor nodes also serve as relays in the routing of the messages to the destination nodes. Each sensor node has the capacity to store messages and decides to transmit or receive messages to or from its immediate neighbors at each time instant, but not both. This simplified rule of transmission allows for a relatively simple representation. We illustrate such a network over a two-dimensional domain in Fig.~\ref{fig:sensornetwork}.
The communication is interference-limited because all nodes share
the same wireless channel. We assume a simple collision protocol: a transmission from a transmitter to a neighboring receiver is successful if and only if none of the other neighbors of the receiver is a transmitter, as illustrated in Fig.~\ref{fig:macrule}.
\begin{figure*}[!t]
\normalsize
\setcounter{mytempeqncnt}{\value{equation}}
\begin{align}
X_N(k + 1, n) - X_N(k, n)=\left\{ \begin{array}{c}
1+G(k,n), \mbox{ with probability}\\
(1-W(n,X_N(k,n)/M))\\
{}\times[P_r(n-1)W(n-1,X_N(k,n-1)/M)(1-W(n+1,X_N(k,n+1)/M))\\
{}+P_l(n+1)W(n+1,X_N(k,n+1)/M)(1-W(n-1,X_N(k,n-1)/M))];\\
{}-1+G(k,n), \mbox{ with probability}\\
W(n,X_N(k,n)/M)\\
{}\times[P_r(n)(1-W(n+1,X_N(k,n+1)/M))(1-W(n+2,X_N(k,n+2)/M)) \\
{}+P_l(n)(1-W(n-1,X_N(k,n-1)/M))(1-W(n-2,X_N(k,n-2)/M))];\\
G(k,n), \mbox{ otherwise}.\end{array} \right.
\label{equ:networktranrule}
\end{align}
\setcounter{equation}{\value{mytempeqncnt}}
\hrulefill
\vspace*{4pt}
\end{figure*}
\addtocounter{equation}{1}
\subsection{Continuum Model in One Dimension}
We first consider the case of a one-dimensional network, where $N$ sensor nodes are uniformly placed over a domain $\mathcal D\subset\real$ and labeled by $n=1,\ldots,N$. The destination nodes
are located on the boundary of $\mathcal{D}$, labeled $n=0$ and $n=N+1$. Again let $ds_N$ be the distance between neighboring nodes.
Let $X_N(k,n)$ in~(\ref{def:MC}) be the queue length of node $n$ at time $k$. Let $M$ in~(\ref{equ:difference}) be the maximum queue length of each node.

\begin{figure}
\centering
\includegraphics[width=70mm]{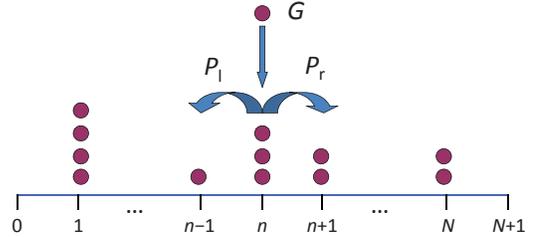}
\caption{An illustration of the time evolution of the queues in the
one-dimensional network model.}
\label{fig:networkrule_conv}
\end{figure}

At each time instant $k=0,1,\ldots$\,, node $n$ decides to be a
transmitter with probability $W(n, X_N(k,n)/M)$. Assume that node $n$
randomly chooses to transmit to the right or the left immediate
neighbor with probability $P_r(n)$ and $P_l(n)$, respectively.
Define $G(k)=[G(k,1),\ldots, G(k,N)]^T$, where $G(k,n)$ is the number
of messages generated at node $n$ at time $k$. We model
$G(k,n)$ by independent Poisson random variables with mean $g(n)$.
The destination nodes at the boundaries of the domain do not have
queues; they simply receive any message transmitted to it and never
itself transmits anything.
We illustrate the time evolution of the queues
in the network in Fig.~\ref{fig:networkrule_conv}.

The sequence $X_N(k)$ defined above forms a Markov chain whose evolution is described by~(\ref{equ:difference}).
According to the behavior of the nodes, the
$n$th component of $F_N(X_N(k)/M,U(k))$, where $n=1,\ldots,N$, is defined
by~(\ref{equ:networktranrule}) at the top of the page, where $X_N(k,n)$ with $n\leq0$ or $n\geq N+1$ are defined to be zero. For
simplicity, in the following parts we set $W(n,X_N(k,n)/M)=X_N(k,n)/M$,
which corresponds to the transmission rule that a node transmits a
message with a probability proportional to its queue length. With
this simplification, for $x=[x_1,\ldots,x_N]^T$, the $n$th component of $F_N(x,U(k))$, where $n=1,\ldots,N$,  is
\begin{align*}
\left\{ \begin{array}{c}
1+G(k,n), \mbox{ with probability}\\
(1-x_n)[P_r(n-1)x_{n-1}(1-x_{n+1})\\
{}+P_l(n+1)x_{n+1}(1-x_{n-1})];\\
{}-1+G(k,n), \mbox{ with probability}\\
x_n[P_r(n)(1-x_{n+1})(1-x_{n+2}) \\
{}+P_l(n)(1-x_{n-1})(1-x_{n-2})];\\
G(k,n), \mbox{ otherwise},\end{array} \right.
\end{align*}
where $x_n$ with $n\leq0$ or $n\geq N+1$ are defined to be zero.

Define $f_N$ as in~(\ref{equ:f_N}). It follows that for $x=[x_1,\ldots,x_N]^T$, the
$n$th component of $f_N(x)$, where $n=1,\ldots,N$,  is
\begin{align}
&(1-x_n)[P_r(n-1)x_{n-1}(1-x_{n+1})\nn\\
&{}+P_l(n+1)x_{n+1}(1-x_{n-1})]\nonumber \\
&{}-x_n[P_r(n)(1-x_{n+1})(1-x_{n+2})\nonumber \\
&{}+P_l(n)(1-x_{n-1})(1-x_{n-2})]+g(n),  \label{equ:networkdifferenceequnew0}
\end{align}
where $x_n$ with $n\leq0$ or $n\geq N+1$ are defined to be zero.
Define the deterministic sequence
$x_N(k)$ as in (\ref{equ:differencenonrand}).

Set $\delta_N$, defined in Section~\ref{subsec:pdeconvergence}, to be $ds_N^2$. Let
\beq
dt_N=\delta_N/M=ds_N^2/M.\label{equ:dtds}
\eeq

Assume
\beq
P_l(n)=p_l(v_N(n))\mbox{ and }
P_r(n)=p_r(v_N(n)),\label{equ:def_Pl_Pr}
\eeq
where $p_l(s)$ and $p_r(s)$ are real-valued functions defined on $\mathcal D$.
As in Section~\ref{sec:randomwalk} we again assume
\beq
p_l(s)=b(s)+c_l(s)ds_N\mbox{ and }
p_r(s)=b(s)+c_r(s)ds_N.
\label{equ:def_pl_pr}
\eeq
Let $c=c_l-c_r$.
Again we call $b$ the diffusion and $c$ the convection.
In order to guarantee that the number of messages entering the system from outside over finite time intervals remains finite throughout the limiting process, we set $g(n)=Mg_p(v_N(n))dt_N$.
Assume $b,c_l,c_r,$ and $g_p$ are in $\mathcal C^1$. Then $f_N\in\mathcal C^1$.

Let $f_N(y_N,s_N)$ be defined as in Section~\ref{subsec:pdeconvergence}.
Then we have the $f$ in~(\ref{equ:w_diff1}):
\begin{align}\label{equ:networkfuncinterior}
f&=b(s)\frac{d}{d s}\left((1-z(s))(1+3z(s))z_s(s)\right)\nn\\
&\quad+2(1-z(s))z_s(s)b_s(s)+z(s)(1-z(s))^2b_{ss}(s)\nn\\
&\quad+\frac{d}{d s}(c(s)z(s)(1-z(s))^2)+g_p(s).
\end{align}
Here, recall that, a single subscript $s$ represents first derivative and a double subscript $ss$ represents second derivative.

Based on the behavior of nodes $n=1$ and $n=N$ next to the destination nodes, we derive the boundary condition for the PDE.
For example, the node $n=1$ receives messages only from the right and encounters no interference when transmitting to the left. Replacing $x_n$ with $n\leq0$ or $n\geq N+1$ by 0 in (\ref{equ:networkdifferenceequnew0}), it follows that the $1$st component of $f_N(x)$ is
\begin{align}
&(1-x_n)P_l(n+1)x_{n+1}\nn\\
&{}-x_n[P_l(n)+P_r(n)(1-x_{n+1})(1-x_{n+2})]+g(n).\label{equ:networkdifferenceequsink1}
\end{align}
Similarly, the $N$th component of $f_N(x)$ is
\begin{align}
&(1-x_n)P_r(n-1)x_{n-1}\nn\\
&{}-x_n[P_r(n)+P_l(n)(1-x_{n-1})(1-x_{n-2})]+g(n).\label{equ:networkdifferenceequsinkN}
\end{align}
Set $\beta_N$, defined in Section~\ref{subsec:pdeconvergence}, to be 1.
Then we have the $h$ in~(\ref{equ:w_diff2}): 
\beq
h = - b(s)z(s)^3 +b(s)z(s)^2 - b(s)z(s).\label{equ:networkfuncbndry}
\eeq
Solving $h=0$ for real $z$, we have the boundary condition $z(t,s)=0$.
This equation might seem confusing to some readers as the limit of (\ref{equ:networkdifferenceequsink1}) and (\ref{equ:networkdifferenceequsinkN}), if it has not been noticed that, unlike $f$, $g$ is the limit of a different function $f_N(y_N,s_N)/\beta_N$.

For fixed $T$, let $z:[0,T]\times \mathcal D \gt\real$ be the solution of the PDE~(\ref{equ:z_pde1}), with boundary condition $z(t,s)=0$ and initial condition $z(0,s)=z_0(s)$, where the right hand side of~(\ref{equ:z_pde1}) is
\begin{align}
&b(s)\frac{\pl}{\pl s}\left((1-z(t,s))(1+3z(t,s))z_s(t,s)\right)\nn\\
&{}+2(1-z(t,s))z_s(t,s)b_s(s)+z(t,s)(1-z(t,s))^2b_{ss}(s)\nn\\
&{}+\frac{\pl}{\pl s}(c(s)z(t,s)(1-z(t,s))^2)+g_p(s).
\label{equ:networkpdeinterior}
\end{align}

In the following we show the convergence of the Markov chain $X_N(k)$ to the PDE solution $z$ to in the one-dimensional network case.
Define $K_N$, $z_N$, $u_N$, and $\ep_N$ as in Section~\ref{subsec:pdeconvergence}.
Throughout this section we assume~(\ref{equ:ep0_0}) holds.
By~(\ref{equ:networkdifferenceequnew0}) and (\ref{equ:networkpdeinterior}), it follows that there exists $0<c<\infty$ such that for $N$ sufficiently large,
\beq
\|\gamma_N \|_\infty^{(N)}<cds_N.\label{inequ:network_u_N_ds_N}
\eeq
Albeit arduous, the algebraic manipulation in getting~(\ref{equ:networkfuncinterior}), (\ref{equ:networkfuncbndry}), and (\ref{inequ:network_u_N_ds_N}) amounts only to algebraic exercises, the concept of which is no more sophisticated than that in getting~(\ref{equ:randomwalkode1}) in Section~\ref{sec:randomwalk}. In practice, we accomplishe such manipulation using symbolic tools provided by computer programs such as Matlab.

By~(\ref{equ:error_sys_H}), for each $N$, for $k=1,\ldots,K_N$ and $n=1,\ldots,N$, we can write $\ep_N(k,n)=H_N^{(k,n)}(u_N)$,
where $H_N^{(k,n)}$ is a real-valued function defined on $\real^{N\times K_N}$.
It follows that $H_N^{(k,n)}(0)=0$ and $H_N^{(k,n)}\in\mathcal C^1$.

Define
\beq
DH_N=\max_{{k=1,\ldots,K_N}\atop{n=1,\ldots,N}}\sum_{{i=1,\ldots,K_N}\atop{j=1,\ldots,N}}\left|\frac{\pl H_N^{(k,n)}}{\pl u(i,j)}(0)\right|,\nn
\eeq
where 0 is in $\real^{N\times K_N}$.

\begin{lemma}\label{lmm:hH}
We have that for each $N$, 
\beq
\mu_N\leq DH_N.\nn 
\eeq
\end{lemma}
\bproof
For each $N$, we have
\begin{align*}
&\max_{{k=1,\ldots,K_N}\atop{n=1,\ldots,N}}\left|\sum_{{i=1,\ldots,K_N}\atop{j=1,\ldots,N}}
\frac{\pl H_N^{(k,n)}}{\pl u(i,j)}(0) u(i,j)\right|\\
&\leq
\max_{{k=1,\ldots,K_N}\atop{n=1,\ldots,N}}\left(\sum_{{i=1,\ldots,K_N}\atop{j=1,\ldots,N}}
\left|\frac{\pl H_N^{(k,n)}}{\pl u(i,j)}(0)\right| |u(i,j)|\right)\\
&\leq DH_N\max_{{i=1,\ldots,K_N}\atop{j=1,\ldots,N}}|u(i,j)|\\
&=DH_N\|u \|_\infty^{(N)}.
\end{align*}
Thus, for each $N$, for all $u\neq0$,
\begin{align}
DH_N\geq\frac{\max_{{k=1,\ldots,K_N}\atop{n=1,\ldots,N}}\left|\sum_{{i=1,\ldots,K_N}\atop{j=1,\ldots,N}}
\frac{\pl H_N^{(k,n)}}{\pl u(i,j)}(0) u(i,j)\right|}{\|u \|_\infty^{(N)}}.\label{inequ:H}
\end{align}

For each $N$, let $v=[v(1),\ldots,v(K_N)]$, where $v(k)=[v(k,1),\dots,v(k,N)]^T$, where for $k=1,\ldots,K_N$ and $n=1,\ldots,N$,
\begin{align*}
v(k,n)=\sgn{\frac{\pl H_N^{(k_0,n_0)}}{\pl u(k,n)}(0)},
\end{align*}
where
\[
(k_0,n_0)\in\argmax_{{k=1,\ldots,K_N}\atop{n=1,\ldots,N}}\sum_{{i=1,\ldots,K_N}\atop{j=1,\ldots,N}}
\left|\frac{\pl H_N^{(k,n)}}{\pl u(i,j)}(0)\right|.
\]
Then
\beq
DH_N=\frac{\max_{{k=1,\ldots,K_N}\atop{n=1,\ldots,N}}\left|\sum_{{i=1,\ldots,K_N}\atop{j=1,\ldots,N}}
\frac{\pl H_N^{(k,n)}}{\pl u(i,j)}(0) v(i,j)\right|}{\|v \|_\infty^{(N)}}.\nn
\eeq
By this and (\ref{inequ:H}) we have
\beq
DH_N=\sup_{u\neq0}\frac{\max_{{k=1,\ldots,K_N}\atop{n=1,\ldots,N}}\left|\sum_{{i=1,\ldots,K_N}\atop{j=1,\ldots,N}}
\frac{\pl H_N^{(k,n)}}{\pl u(i,j)}(0) u(i,j)\right|}{\|u \|_\infty^{(N)}}.\label{equ:DH_N}
\eeq

By Taylor's theorem, for each $N$, for $k=1,\ldots,K_N$ and $n=1,\ldots,N$, there exists $\tld H_N^{(k,n)}(u)$ such that
\begin{align}
H_N^{(k,n)}(u)=\sum_{{i=1,\ldots,K_N}\atop{j=1,\ldots,N}}
\frac{\pl H_N^{(k,n)}}{\pl u(i,j)}(0) u(i,j)+\tld H_N^{(k,n)}(u),\label{equ:H_N_kn}
\end{align}
and for $i=1,\ldots,K_N$ and $j=1,\ldots,N$,
\begin{align*}
\lim_{u\to0}
\frac{|\tld H_N^{(k,n)}(u)|}{\|u\|_\infty^{(N)}}=0.
\end{align*}
Hence for each $\ep>0$, there exists $\delta$ such that for $\|u\|_\infty^{(N)}<\delta$, we have
\begin{align*}
\frac{|\tld H_N^{(k,n)}(u)|}{\|u\|_\infty^{(N)}}<\ep.
\end{align*}
Then for $\|u\|_\infty^{(N)}\leq\alpha\leq\delta$,
\begin{align*}
\sup_{\|u\|_\infty^{(N)}\leq\alpha}\frac{|\tld H_N^{(k,n)}(u)|}{\|u\|_\infty^{(N)}}<\ep.
\end{align*}
Therefore, for $i=1,\ldots,K_N$ and $j=1,\ldots,N$,
\begin{align}
\lim_{\alpha\to0}
\sup_{\|u\|_\infty^{(N)}\leq\alpha}\frac{|\tld H_N^{(k,n)}(u)|}{\|u\|_\infty^{(N)}}=0.\label{equ:tld_H}
\end{align}
By~(\ref{equ:H_N_kn}), for each $N$,
\begin{align*}
&\|H_N(u)\|_\infty^{(N)}\leq \max_{{k=1,\ldots,K_N}\atop{n=1,\ldots,N}}\left|\tld H_N^{(k,n)}(u)\right|\\
&\quad+\max_{{k=1,\ldots,K_N}\atop{n=1,\ldots,N}}\left|\sum_{{i=1,\ldots,K_N}\atop{j=1,\ldots,N}}
{\pl H_N^{(k,n)}}{\pl u(i,j)}(0) u(i,j)\right|.
\end{align*}
Hence
\begin{align*}
\mu_N&\leq\lim_{\alpha\to0}\sup_{\|u\|_\infty^{(N)}\leq\alpha}\left(\frac{\max_{{k=1,\ldots,K_N}\atop{n=1,\ldots,N}}\left|\tld H_N^{(k,n)}(u)\right|}{\|u \|_\infty^{(N)}}\right.\\
&\left.\quad+\frac{\max_{{k=1,\ldots,K_N}\atop{n=1,\ldots,N}}\left|\sum_{{i=1,\ldots,K_N}\atop{j=1,\ldots,N}}
\frac{\pl H_N^{(k,n)}}{\pl u(i,j)}(0) u(i,j)\right|}{\|u \|_\infty^{(N)}}\right).
\end{align*}
Hence by~(\ref{equ:DH_N}) and (\ref{equ:tld_H}), we finish the proof.
\eproof
Notice that $DH_N$ is essentially the induced $\infty$-norm of the linearized version of the operator $H_N$.

Now we present a lemma on the condition of the sequence $\{\mu_N\}$ being bounded for the one-dimensional network case.
\blemma\label{lmm:g_bnded}
In the one-dimensional network case, assume that the function
\begin{align}
\max\{|z|,|z_{s}|,|z_{ss}|,|b|,|b_{s}|,|b_{ss}|,|c|,|c_{s}|\}\label{inequ:bnded_zbc}
\end{align}
of $(t,s)$ is bounded on $[0,T]\times\mathcal D$.
Then $\{\mu_N\}$ is bounded.
\elemma

\bproof
Define
\[
A_N(k)=I_N+\frac{1}{M}Df_N(z_N(k)),
\]
where $I_N$ be the identity matrix in $\real^{N\times N}$.
It follows from~(\ref{equ:ep1}) that for each $N$ and for $k=0,\ldots,K_N$,
\begin{align}
\ep_N(k+1)=A_N(k)\ep_N(k)+\frac{\tld f_{N}(\ep_N(k))}{M}+dt_Nu_N(k).\nn
\end{align}
It follows that
\begin{align*}
\ep_N(k)&=dt_N(A_N(k-1)\ldots A(1)u_N(0)\nn\\
&\quad+A_N(k-1)\ldots A(2)u_N(1)\nn\\
&\quad+\ldots \nn\\
&\quad+A_N(k-1)u_N(k-2)+u_N(k-1))\nn\\
&\quad+\frac{1}{M}(A_N(k-1)\ldots A(2)\tld f_{N}(\ep_N(1))\nn\\
&\quad+A_N(k-1)\ldots A(3)\tld f_{N}(\ep_N(2))\nn\\
&\quad+\ldots \nn\\
&\quad+A_N(k-1)\tld f_{N}(\ep_N(k-2))\nn\\
&\quad+\tld f_{N}(\ep_N(k-1)).\nn
\end{align*}
Define
\begin{align}
&B_N^{(k,n)}=\left\{
\begin{array}{lr}
0,& 0\leq n<k-3;\\
I_N,& n=k-3;\\
A_N(k-1)\ldots A_N(n+1),& n\geq k-2.\label{def:B_N}
\end{array}
\right.
\end{align}
%
It follows that
\begin{align}
\frac{\pl H_N^{(k,n)}(u)}{\pl u(i,j)}(0)=B_N^{(k,i)}(n,j)dt_N.\nn
\end{align}
Hence by Lemma~\ref{lmm:hH},
\begin{align}
\mu_N&\leq\max_{{k=1,\ldots,K_N}\atop{n=1,\ldots,N}}\sum_{{i=1,\ldots,K_N}\atop{j=1,\ldots,N}}\left|B_N^{(k,i)}(n,j)\right|dt_N.\label{equ:h_B}
\end{align}

By~(\ref{equ:networkdifferenceequnew0}), for fixed $N$, for $x=[x_1,\ldots,x_N]^T$, the $(n,m)$th component of $Df_N(x):=\frac{\pl f_N^{(n)}}{\pl x_m}(x)$, where $n,m=1,\ldots,N$, is
\begin{align*}
&\left\{\begin{array}{lr}
P_l(n)x_n(1-x_{n-1}), & m=n-2;\\
(1-x_n)[P_r(n-1)(1-x_{n+1})\\
\quad-P_l(n+1)x_{n+1}]\\
\quad+P_l(n)x_n(1-x_{n-2}), &  m=n-1;\\
-[P_r(n-1)x_{n-1}(1-x_{n+1})\\
\quad+P_l(n+1)x_{n+1}(1-x_{n-1})]\\
\quad-[P_r(n)(1-x_{n+1})(1-x_{n+2})\\
\quad+P_l(n)(1-x_{n-1})(1-x_{n-2})], &  m=n;\\
(1-x_n)[P_l(n+1)(1-x_{n-1})\\
\quad-P_r(n-1)x_{n-1}]\\
\quad+P_r(n)x_n(1-x_{n+2})],& m=n+1;\\
P_r(n)x_n(1-x_{n+1}), &  m=n+2;\\
0  &\mbox{ other wise,}
\end{array} \right.
\end{align*}
where $x_n$ with $n\leq0$ or $n\geq N+1$ are defined to be zero.

Denote  the  induced $\infty$-norm on $\real^{N\times N}$ again by $\|\cdot\|^{(N)}_\infty$. That is,  for $A\in \real^{N\times N}$, with the $(i,j)$th element being $A(i,j)$,
\[
\|A\|^{(N)}_\infty=\max_{1\leq i\leq N}\sum_{j=1}^N|A(i,j)|,
\]
which is simply the maximum absolute row sum of the matrix.
Then we have,
\begin{align*}
&\|A_N(k)\|^{(N)}_\infty\\
&=\max_{n=1,\ldots,N}\frac{1}{M}(|P_l(n)z_N(k,n)(1-z_N(k,n-1))|\\
&\quad+|(1-z_N(k,n))[P_r(n-1)(1-z_N(k,n+1))\\
&\quad-P_l(n+1)z_N(k,n+1)]\\
&\quad+P_l(n)z_N(k,n)(1-z_N(k,n-2))|\\
&\quad+|M-[P_r(n-1)z_N(k,n-1)(1-z_N(k,n+1))\\
&\quad+P_l(n+1)z_N(k,n+1)(1-z_N(k,n-1))]\\
&\quad-[P_r(n)(1-z_N(k,n+1))(1-z_N(k,n+2))\\
&\quad+P_l(n)(1-z_N(k,n-1))(1-z_N(k,n-2))]|\\
&\quad+|(1-z_N(k,n))[P_l(n+1)(1-z_N(k,n-1))\\
&\quad-P_r(n-1)z_N(k,n-1)]\\
&\quad+P_r(n)z_N(k,n)(1-z_N(k,n+2))]|\\
&\quad+|P_r(n)z_N(k,n)(1-z_N(k,n+1))|).
\end{align*}
Put~(\ref{equ:def_Pl_Pr}), (\ref{equ:def_pl_pr}), and the Taylor's expansions (\ref{equ:taylor_00}), (\ref{equ:taylor_01}), and (\ref{equ:taylor_02}) of $z,b$, and $c$, respectively, into the above equation and rearrange. (Again we omit the detailed algebraic manipulation here.)
Then we have that there exist $0<c_1<\infty$ such that for each $N$, for $k=1,\ldots,K_N$ and $n=1,\ldots,N$,
\begin{align*}
&\|A_N(k)\|^{(N)}_\infty\\
&\leq\max_{n=1,\ldots,N}|-c_{s}(v_N(n)) - b_{ss}(v_N(n)) \\
&\quad- 2b(v_N(n))z_{ss}(t_N(k),v_N(n))\\
&\quad+ 4b_{ss}(v_N(n))z(t_N(k),v_N(n)) \\
&\quad+ 2b_s(v_N(n))z_s(t_N(k),v_N(n))\\
&\quad+ 4c_s(v_N(n))z(t_N(k),v_N(n))\\
&\quad+ 4c(v_N(n))z_s(t_N(k),v_N(n))\\
&\quad+ 6b(v_N(n))z_s(t_N(k),v_N(n))^2 \\
&\quad- 3b_{ss}(v_N(n))z(t_N(k),v_N(n))^2 \\
&\quad- 3c_{s}(v_N(n))z(t_N(k),v_N(n))^2 \\
&\quad+ 6b(v_N(n))z(t_N(k),v_N(n))z_{ss}(t_N(k),v_N(n))\\
&\quad- 6c(v_N(n))z(t_N(k),v_N(n))z_s(t_N(k),v_N(n))|\frac{ds^2}{M}\\
&\quad + c_1\frac{ds^3}{M}+1\\
&:= \max_{n=1,\ldots,N}|q(t_N(k),v_N(n))|\frac{ds^2}{M} + c_1\frac {ds^3}{M}+1.
\end{align*}
Since~(\ref{inequ:bnded_zbc}) is bounded, there exists $0<c_2<\infty$ such that $|q(t,s)|<c_2$ for all $(t,s)\in[0,T]\times\mathcal D$.
Hence for each $N$ and for $k=0,\ldots,K_N$,
\begin{align*}
\|A_N(k)\|^{(N)}_\infty\leq1+c_2\frac{ds_N^2}{M}+c_1\frac{ds_N^3}{M}.
\end{align*}
Hence there exists $0<c_3<\infty$, for $N$ sufficiently large and for $k=0,\ldots,K_N$,
\begin{align*}
\|A_N(k)\|^{(N)}_\infty\leq1+c_3\frac{ds_N^2}{M}=1+c_3dt_N.
\end{align*}

Hence by~(\ref{def:B_N}) and (\ref{equ:h_B}), for $N$ sufficiently large,
\begin{align*}
\mu_N&\leq K_Ndt_N(1+c_3dt_N)^{K_N}.
\end{align*}
Since $T<\infty$, there exist $0<c_4<\infty$ such that for each $N$, $K_Ndt_N<c_4$.
But as $N\to\infty$, $K_N\to\infty$, and
\[
(1+c_3dt_N)^{K_N}=\left(1+\frac{c_3 T }{K_N}\right)^{K_N}\to e^{c_3T}.
\]
Therefore, there exist $0<c_5<\infty$ such that for each $N$, $\mu_N<c_5$.
This finishes the proof.
\eproof

\bprop
In the one-dimensional network case, suppose that the assumption in Lemma~\ref{lmm:g_bnded} holds. Then
\beq
\|X_{N} - z_N \|_\infty^{(N)}= O(ds_N)\mbox{ a.s.\ on }[0,T]\times \mathcal{D}.\nn
\eeq
\eprop
\bproof
By~(\ref{equ:dtds}) and (\ref{inequ:network_u_N_ds_N}), there exists $0<c<\infty$ such that for $N$ sufficiently large,
\[
\max\{\gamma_N,ds_N,dt_N\}\leq cds_N.
\]
One can now easily verify that the assumptions in Theorem~\ref{thm:main_thm} hold.
Then by Theorem~\ref{thm:main_thm} the desired result holds.
\eproof
This proposition states that in the one-dimensional network case, as $M \gt \infty$ and $N \gt \infty$, $X_{N}$ converges uniformly to $z_N$ a.s., and at least with the same rate as $ds_N$.
Analogously, for the continuous-time-space extension $X_{pN}$ of $X_N(k)$, given the same assumption as in the above theorem, by Theorem~\ref{thm:oldmainthm}, we have
\beq
\|X_{pN} - z \|_\infty^{(p)}= O(ds_N)\mbox{ a.s.\ on }[0,T]\times \mathcal{D}.\nn
\eeq


\subsubsection{Interpretation of the approximation PDE}
\begin{figure}
\centering
\includegraphics[width=80mm]{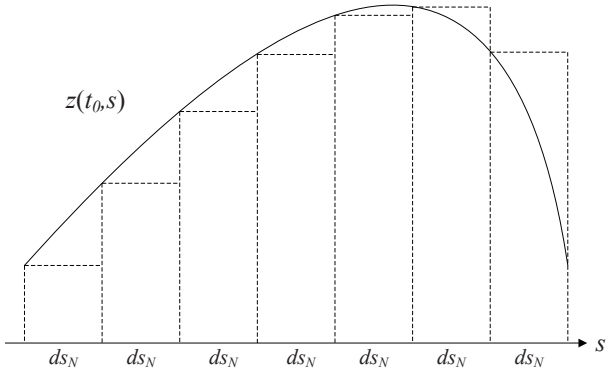}\\
\caption{The PDE solution $z(t,s)$, at $t=t_o$ approximating the normalized queue lengths of a one-dimensional network.}
\label{fig:PDE_physical_meaning}
\end{figure}

Now we make some remarks on how to use a given approximating PDE.
First, for fixed $N$ and $M$, the normalized queue length of node $n$ at time $k$, is approximated by the value of the PDE solution $z$ at the corresponding point in $[0,T]\times\mathcal D$, i.e.,
\[
z((t_N(k),v_N(n)))\approx \frac{X_N(k,n)}{M}.
\]
Second, we show how to interpret
\[
C(t_o):=\int_{\mathcal D} z(t_o,s)ds_N,
\]
the area below the curve $z(t_o,s)$ for fixed $t_o\in[0,T]$.
Let $k_o=\floor{t_o/dt_N}$.
Then we have
\[
z(t_o,v_N(n))ds_N\approx \frac{X_N(k_o,n)}{M}ds_N,
\]
the area of the $n$th rectangle in Fig.~\ref{fig:PDE_physical_meaning}.
Hence
\[
C(t_o)\approx\sum_{n=1}^{N}z(t_o,v_N(n))ds_N\approx\sum_{n=1}^{N}\frac{X_N(k_o,n)}{M}ds_N,
\]
the sum of all rectangles.
If we assume that all messages in the queue have roughly the same bits, and think of $ds_N$ as the ``coverage'' of each node, then the area under any segment of the curve measures a kind of ``data-coverage product'' of the nodes covered by the segment, in the unit of ``bit$\cdot$meter''.
As $N\to\infty$, the total normalized queue length $\sum_{n=1}^{N}X_N(k_o,n)/M$ of the network does go to infinity; however, the coverage $ds_N$ of each node goes to 0. Hence the sum of the ``data-coverage product'' can be approximated by the finite area $C(t_o)$.

\subsubsection{Comparison between PDE approximation and Monte Carlo simulation: One dimension}
We compare the PDE approximation obtained from our approach with the Monte Carlo simulations for a network over the domain $\mathcal D=[-1,1]$. We use the initial condition $z_0(s)=l_1e^{-s^2}$, where $l_1>0$ is a constant, so that initially the nodes in the middle have messages to transmit, while those near the boundaries have very few. We set the message generation rate $g_p(s)=l_2e^{-s^2}$, where $l_2 > 0$ is a parameter determining the total load of the system.

\begin{figure}
\centering
\includegraphics[width=80mm]{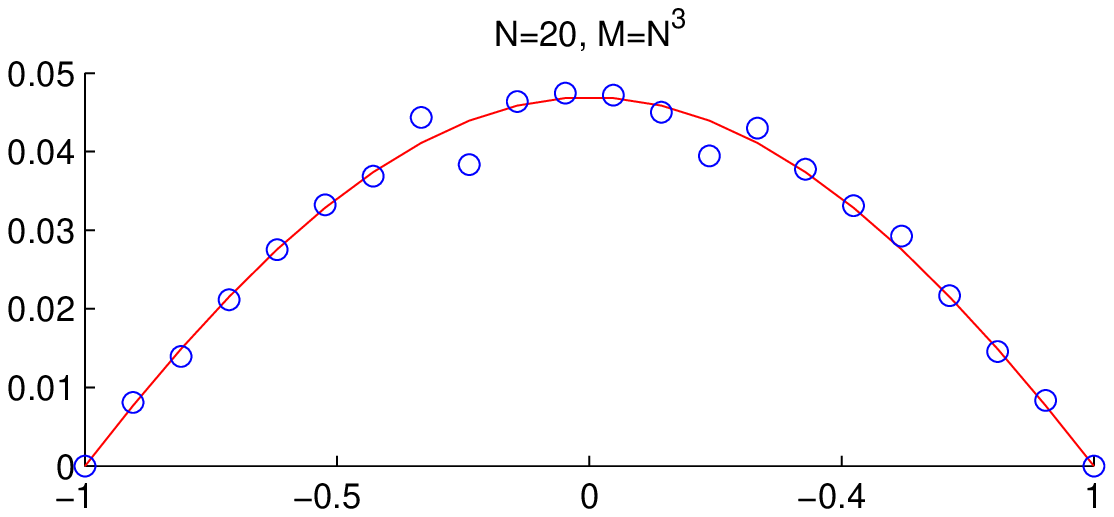}\\
\includegraphics[width=80mm]{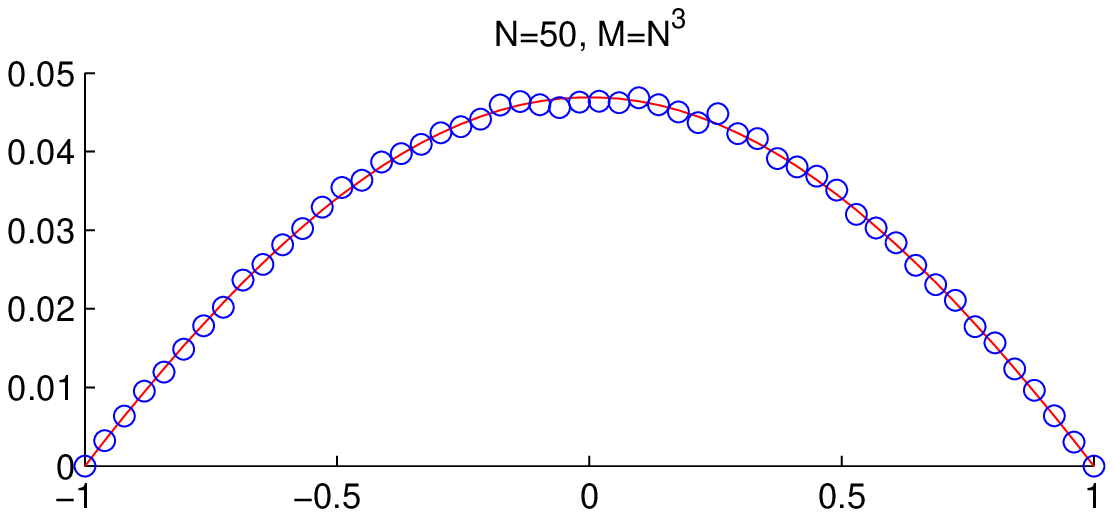}\\
\includegraphics[width=80mm]{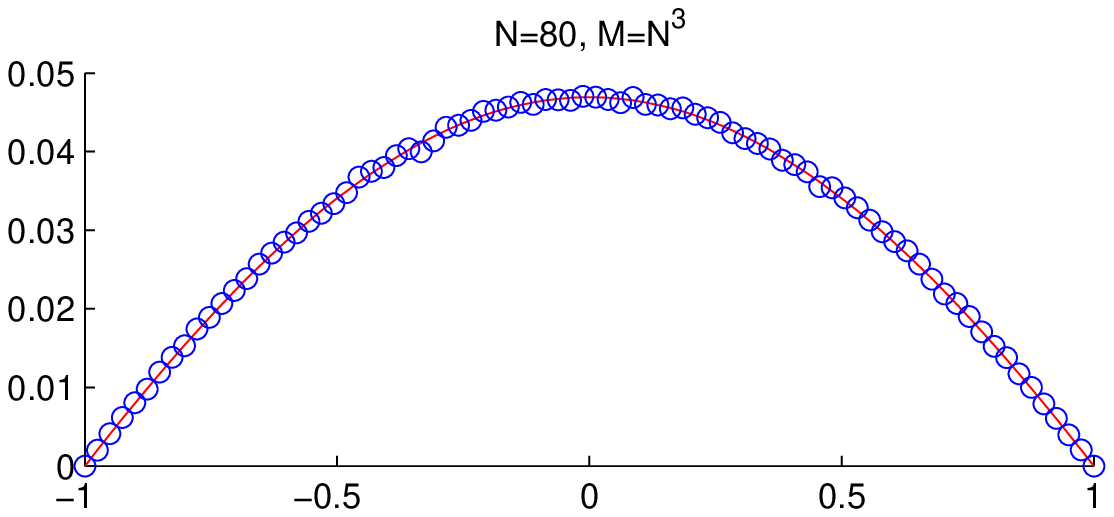}\\
\scriptsize{$\circ$ Monte Carlo simulation  \hspace{5mm} ------ PDE solution}
\caption{The Monte Carlo simulations (with different $N$ and $M$) and the PDE solution of a one-dimensional network, with $b=1/2$ and $c=0$, at $t=1s$.}
\label{fig:simulationresult_nonconv}
\end{figure}

\begin{figure}
\centering
\includegraphics[width=80mm]{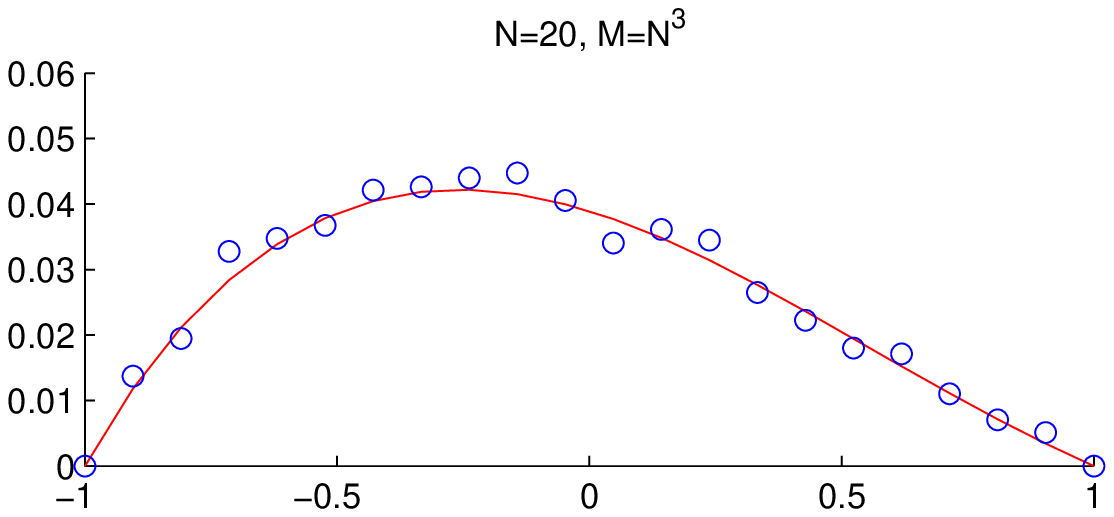}\\
\includegraphics[width=80mm]{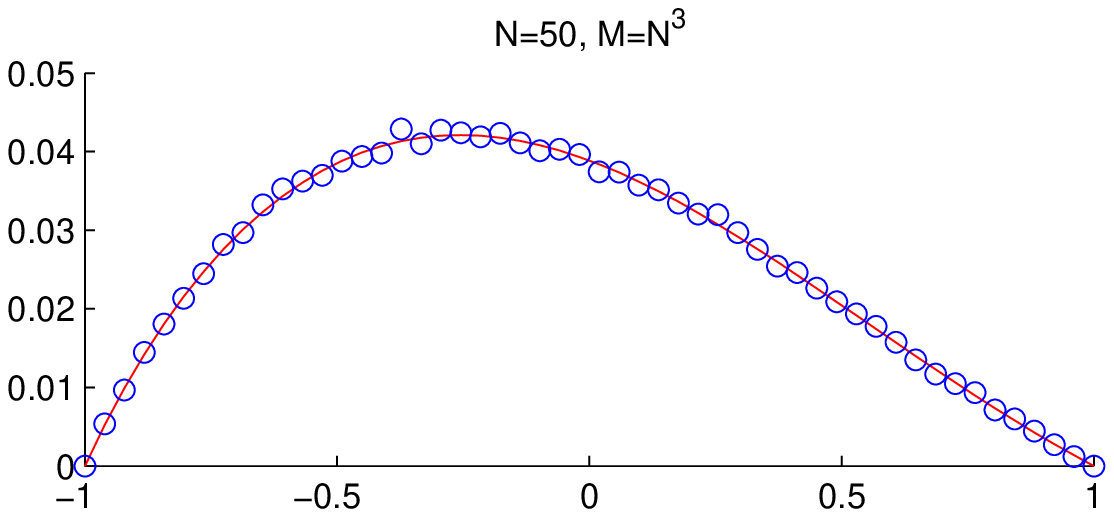}\\
\includegraphics[width=80mm]{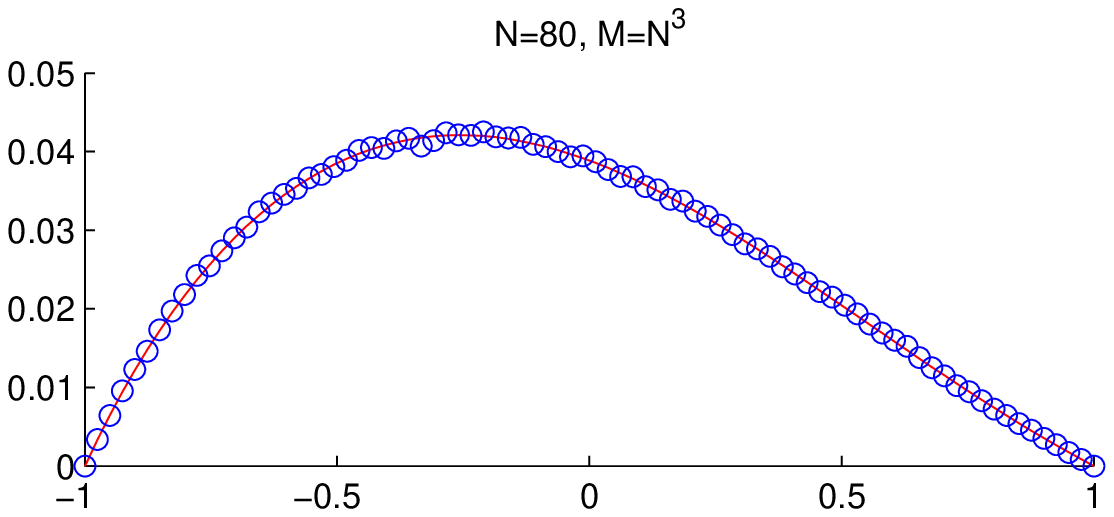}\\
\scriptsize{$\circ$ Monte Carlo simulation  \hspace{5mm} ------ PDE solution}
\caption{The Monte Carlo simulations (with different $N$ and $M$) and the PDE solution of a one-dimensional network, with $b=1/2$ and $c=1$, at $t=1s$.}
\label{fig:simulationresult_conv}
\end{figure}

We use three sets of values of $N=20,50,80$ and $M=N^3$, and show the PDE solution and the  Monte Carlo simulation results with different $N$ and $M$ at $t=1s$.
The networks have diffusion coefficient $b=1/2$  and convection coefficient $c=0$ in Fig.~\ref{fig:simulationresult_nonconv} and $c=1$ in Fig.~\ref{fig:simulationresult_conv}, respectively, where the x-axis denotes the node location and y-axis denotes the normalized queue length.


For the three sets of the values of $N=20,50,80$ and $M=N^3$ and with $c=0$, the maximum absolute errors of the PDE approximation are $5.6\times10^{-3}$, $1.3\times10^{-3}$, and $1.1\times10^{-3}$, respectively; and with $c=1$, the errors are $4.4\times10^{-3}$, $1.5\times10^{-3}$, and $1.1\times10^{-3}$, respectively. As we can see, as $N$ and $M$ increase, the resemblance between the Monte Carlo simulations and the PDE solution becomes stronger. In the case of very large $N$ and $M$, it is difficult to distinguish the results.

We stress that the PDEs only took fractions of a second to solve on a computer, while the Monte Carlo simulations took time on the order of tens of hours. We could not do Monte Carlo simulations of any larger networks because of prohibitively long computation time.
\subsection{Continuum Model in Two Dimensions}
Generalization of the continuum model to higher dimensions is straightforward, except for more arduous algebraic manipulation.
Now we consider the two-dimensional network of $N_1\times N_2$ sensor nodes.
The nodes are uniformly placed over the domain $\mathcal D\subset\real^2$ and labeled by $(n,m)$, where $n=1,\ldots,N_1$ and $m=1,\ldots,N_2$. Again let the distance between neighboring nodes be $ds_N$. Assume that the node at location $(n,m)$ randomly chooses to transmit to the north, east, south, or west immediate neighbor with probabilities $P_{e}(n,m)=b_1(s)+c_e(s)ds_N$, $P_{w}(n,m)=b_1(s)+c_w(s)ds_N$, $P_{n}(n,m)=b_2(s)+c_n(s)ds_N$, and $P_{s}(n,m)=b_2(s)+c_s(s)ds_N$, respectively. Define $c_1=c_w-c_e$ and $c_2=c_s-c_n$.

The derivation of the approximating PDE is similar to those of the one-dimensional cases, except that we now have to consider transmission to and interference from four directions instead of two.
We present the approximating PDE here without the detailed derivation:
\begin{align*}
\dot z&=\sum_{j=1}^2b_j\frac{\pl}{\pl s_j}\left((1+5z)(1-z)^{3}\frac{\pl z}{\pl s_j}\right) +2(1-z)^{3}\frac{\pl z}{\pl s_j}\nn\\
&\quad{}\times\frac{d b_j}{d s_j}+z(1-z)^{4}\frac{d^2 b_j}{d s_j^2}+\frac{\pl}{\pl s_j}\left(c_jz(1-z)^{4}\right),
\end{align*}
with boundary condition $z(t,s)=0$ and initial condition $z(0,s)=z_0(s)$, where $t\in[0,T]$ and $s=(s_1,s_2)\in\mathcal D.$

\subsubsection{Comparison between PDE approximation and Monte Carlo simulations: Two dimensions}
We compare the PDE approximation and the Monte Carlo simulations of a network over the domain $D=[-1,1]\times[-1,1]$. We use the initial condition $z_0(s)=l_1e^{-(s_1^2+s_2^2)}$, where $l_1>0$ is a constant, so that initially the nodes in the center have messages to transmit, while those near the boundary have very few. We set the message generation rate $g_p(s)=l_2e^{-(s_1^2+s_2^2)}$, where $l_2 > 0$ is a parameter determining the total load of the system.

\begin{figure}
\centering
\includegraphics[width=80mm]{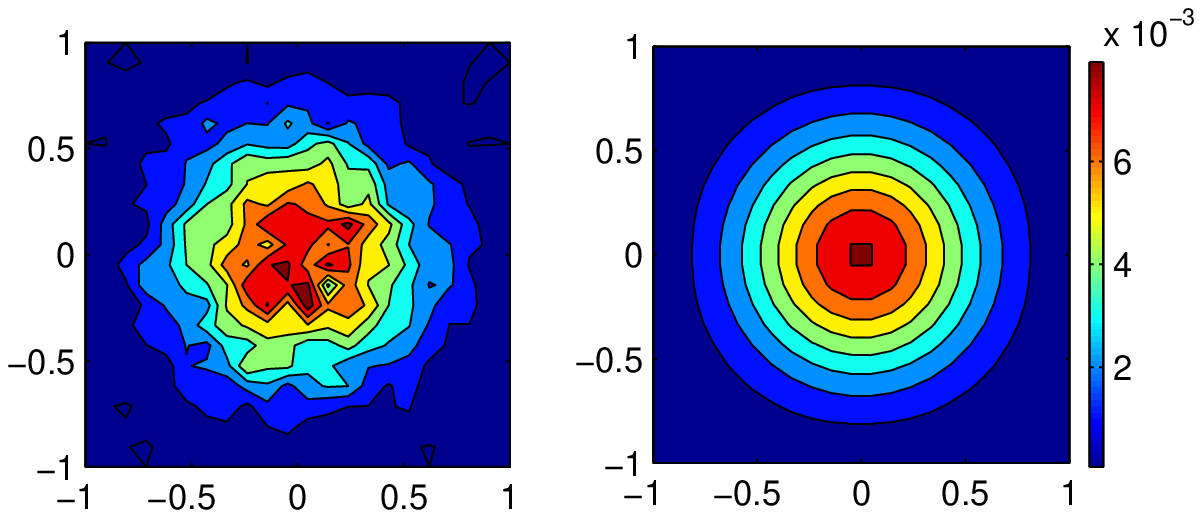}\\
\includegraphics[width=80mm]{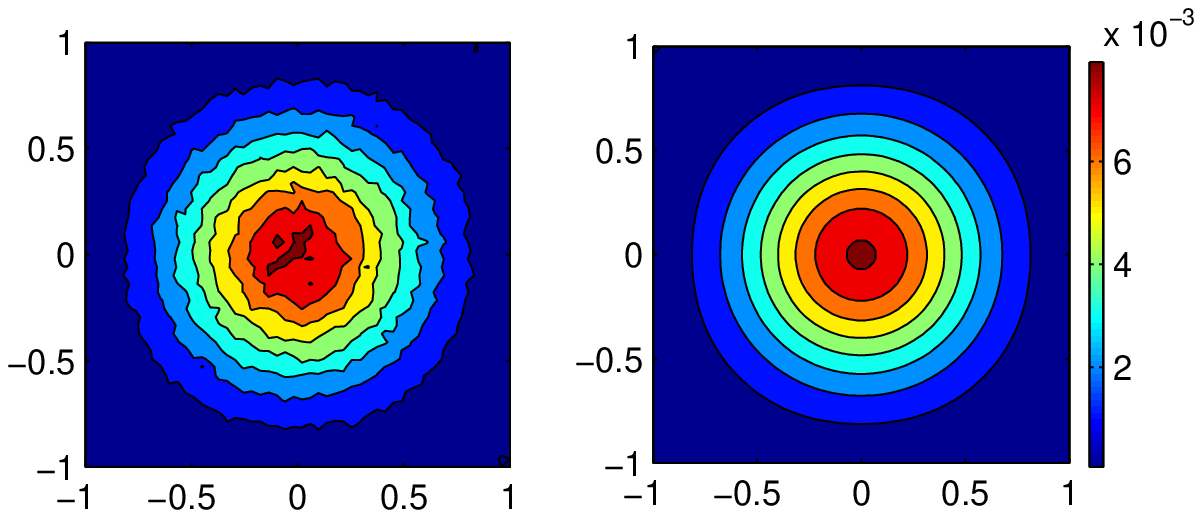}\\
\includegraphics[width=80mm]{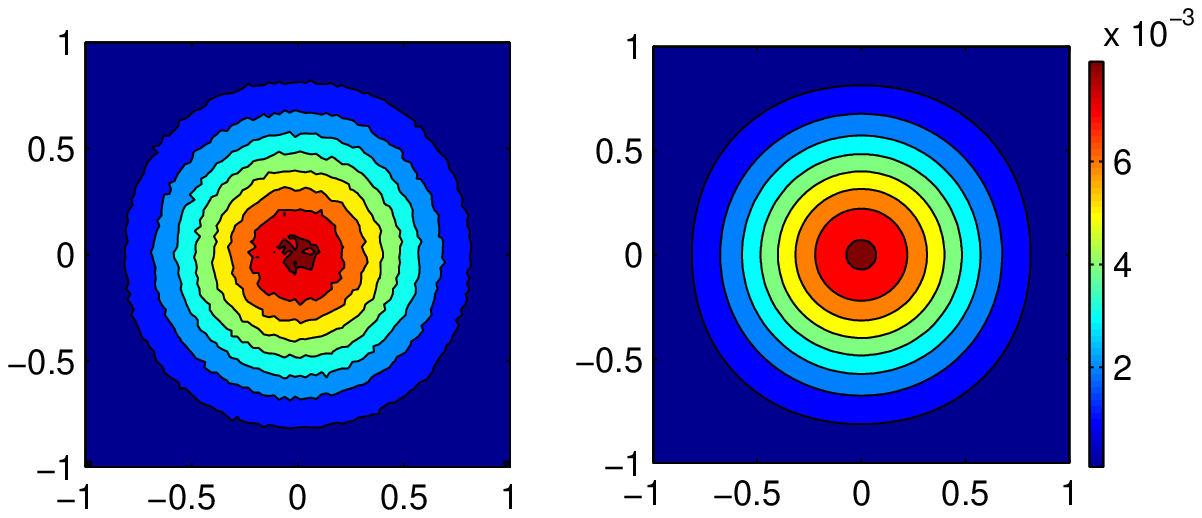}\\
\scriptsize{\hspace{-2mm}Monte Carlo simulations  \hspace{16mm}  PDE solution}
\caption{The Monte Carlo simulations (from top to bottom, with $N_1=N_2=20,50,80$, respectively, and $M=N_1^3$) and the PDE solution of a two-dimensional network, with $b_1=b_2=1/4$ and $c_1=c_2=0$, at $t=0.1s$.}
\label{fig:simulationresult_2D_nonconv}
\end{figure}

\begin{figure}[h]
\centering
\includegraphics[width=80mm]{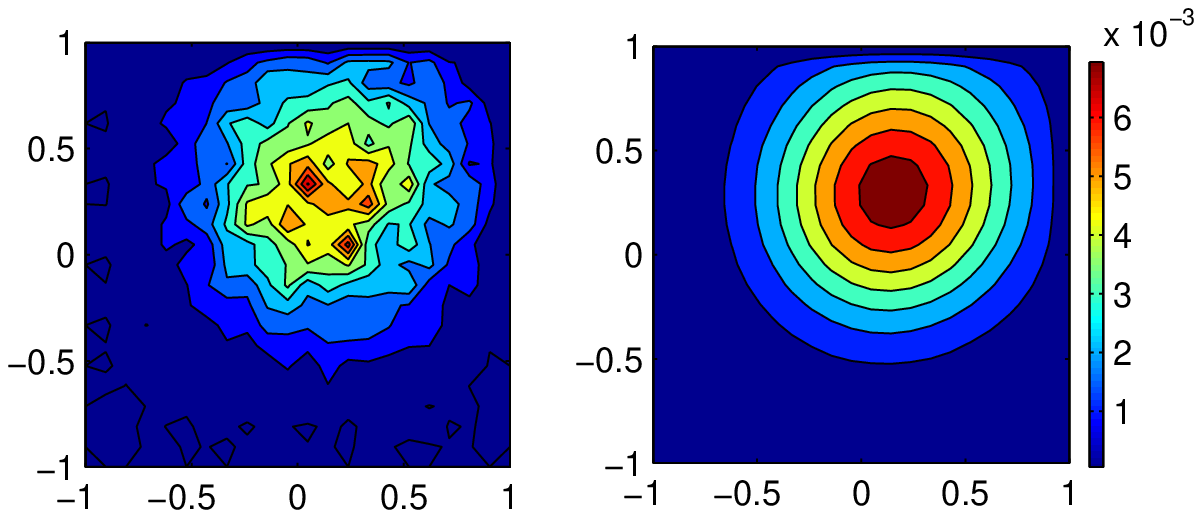}\\
\includegraphics[width=80mm]{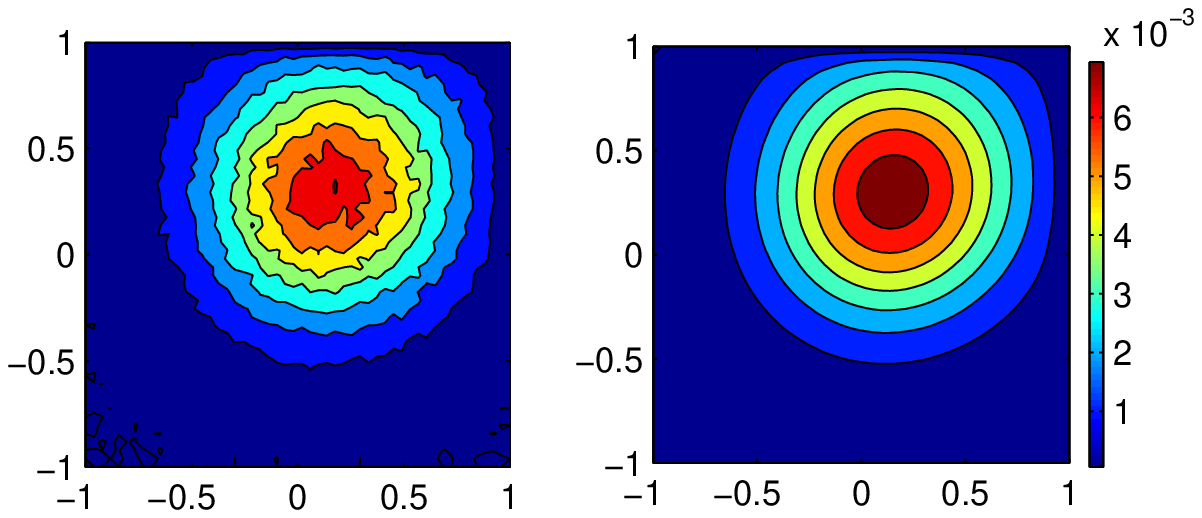}\\
\includegraphics[width=80mm]{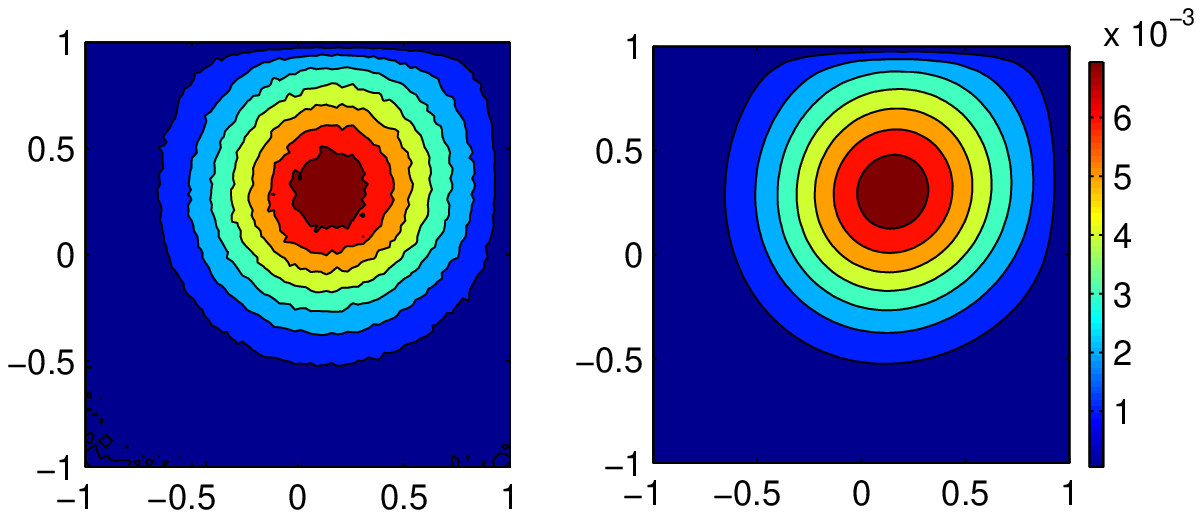}\\
\scriptsize{\hspace{-2mm}Monte Carlo simulations  \hspace{16mm}  PDE solution}
\caption{The Monte Carlo simulations (from top to bottom, with $N_1=N_2=20,50,80$, respectively, and $M=N_1^3$) and the PDE solution of a two-dimensional network, with $b_1=b_2=1/4$ and $c_1=-2,c_2=-4$, at $t=0.1s$.}
\label{fig:simulationresult_2D_conv}
\end{figure}

We use three different sets of the values of $N_1\times N_2$ and $M$, where $N_1=N_2=20,50,80$ and $M=N_1^3$.
We show the contours of the normalized queue length from the PDE solution and the Monte Carlo simulation results with different sets of values of $N_1$, $N_2$, and $M$, at $t=0.1s$.
The networks have diffusion coefficients $b_1=b_2=1/4$ and convection coefficients $c_1=c_2=0$ and $c_1=-2,c_2=-4$ in Fig.~\ref{fig:simulationresult_2D_nonconv} and Fig.~\ref{fig:simulationresult_2D_conv}, respectively.
It took 3 days to do the Monte Carlo simulation of the network at $t=0.1s$ with $80\times80$ nodes and the maximum queue length $M=80^3$, while the PDE solved on the same machines took less than a second. We could not do Monte Carlo simulations of any larger networks or greater values of $t$.

For the three sets of values of $N_1=N_2=20,50,80$ and $M=N_1^3$ and with $c_1=c_2=0$, the maximum absolute errors are $3.2\times10^{-3}$, $1.1\times10^{-3}$, and $6.8\times10^{-4}$, respectively; and with $c_1=-2,c_2=-4$, the errors are $4.1\times10^{-3}$, $1.0\times10^{-3}$, and $6.6\times10^{-4}$, respectively. Again the accuracy of the continuum model increases with $N_1$, $N_2$, and $M$.

\section{Conclusion and Future Work}
In this paper we analyze the convergence of a sequence of Markov
chains to its continuum limit, the solution of a PDE, in a two-step procedure.
We provide precise sufficient conditions for the convergence and the explicit rate of the convergence. Based on such convergence we approximate the Markov chain modeling a large wireless sensor network by a nonlinear diffusion-convection PDE.

With the sophisticated mathematical tools available for PDEs, this approach provides a framework to model and simulate networks with a very large number of components, which is practically infeasible for Monte Carlo simulation.
Such a tool enables us to tackle problems such as performance analysis and prototyping, resource provisioning, network design, network parametric optimization, network control, network tomography, and inverse problems, for very large networks. For example, we can now use the PDE model to optimize some performance metric of a large network by adjusting the placement of destination nodes or the routing parameters (coefficients in convection terms), with relatively negligible computational overhead compared with that of the same task done by Monte Carlo simulation.

The approximation approach can be extended in future work with more specific considerations regarding the network, which can significantly affect the derivation of the continuum model.
For example, we can seek to establish continuum models for other domains such as the Internet, cellular networks, and traffic networks; we can consider more boundary conditions other than sinks, including walls, semi-permeating walls, and their composition; the nodes could be nonuniformly located, even mobile; transmission could happen between nodes that are not immediate neighbors; and the interference between nodes could behave differently in the presence of power control.

\bibliography{ieeeabrv,bibfilezhangy}
\bibliographystyle{ieeetran}

\end{document}

%% file: pream.tex
\usepackage{amsmath}
\usepackage{amssymb}
\usepackage{amsbsy}

\newcommand{\floor}[1]{\lfloor #1\rfloor}

\newcommand{\tld}{\tilde}
\newcommand{\pl}{\partial}

\newcommand{\ep}{\varepsilon}

\newcommand{\argmax}{\mathop{\mbox{\rm arg\,max}}}

\newcommand{\sgn}{\mathop{\mbox{\rm sgn}}}

\newcommand{\gt}{\rightarrow}

\newcommand{\blackbox}{\vrule height7pt width5pt depth1pt}

\newcommand{\bit}{\begin{itemize}}
\newcommand{\eit}{\end{itemize}}
\newcommand{\ben}{\begin{enumerate}}
\newcommand{\een}{\end{enumerate}}
\newcommand{\bdesc}{\begin{description}}
\newcommand{\edesc}{\end{description}}
\newcommand{\beq}{\begin{equation}}
\newcommand{\eeq}{\end{equation}}
\newcommand{\beqarr}{\begin{eqnarray}}
\newcommand{\eeqarr}{\end{eqnarray}}
\newcommand{\beqarrn}{\begin{eqnarray*}}
\newcommand{\eeqarrn}{\end{eqnarray*}}
\newcommand{\nn}{\nonumber}
\newtheorem{thm}{Theorem}
\newcommand{\bthm}{\begin{thm}}
\newcommand{\ethm}{\end{thm}}
\newtheorem{prop}{Proposition}
\newcommand{\bprop}{\begin{prop}}
\newcommand{\eprop}{\end{prop}}
\newcommand{\bproof}{\begin{proof}}
\newcommand{\eproof}{\end{proof}}
\newenvironment{proofof}[1]{\begin{trivlist}\item[]{\emph{Proof of #1:}\hspace{2mm} }}{\hfill\blackbox\end{trivlist}}
\newcommand{\bproofof}{\begin{proofof}}
\newcommand{\eproofof}{\end{proofof}}
\newenvironment{rem}{\begin{trivlist}\item[]{\bf Remark:}\hspace{4mm}}{\end{trivlist}}
\newcommand{\brem}{\begin{rem}}
\newcommand{\erem}{\end{rem}}
\newenvironment{rems}{\begin{trivlist}\item[]{\bf Remarks}\begin{itemize}}{\end{itemize}\end{trivlist}}
\newcommand{\brems}{\begin{rems}}
\newcommand{\erems}{\end{rems}}
\newtheorem{assmp}{Assumption}
\newcommand{\bassmp}{\begin{assmp}}
\newcommand{\eassmp}{\end{assmp}}
\newtheorem{lemma}{Lemma}
\newcommand{\blemma}{\begin{lemma}}
\newcommand{\elemma}{\end{lemma}}
\newtheorem{cor}{Corollary}
\newcommand{\bcor}{\begin{cor}}
\newcommand{\ecor}{\end{cor}}
\newtheorem{fact}{Fact}
\newcommand{\bfact}{\begin{fact}}
\newcommand{\efact}{\end{fact}}
\newtheorem{examp}{Example}
\newcommand{\bexamp}{\begin{examp}\rm}
\newcommand{\eexamp}{\end{examp}}
\newtheorem{defn}{Definition}
\newcommand{\bdefn}{\begin{defn}\rm}
\newcommand{\edefn}{\end{defn}}
\newtheorem{prob}{Problem}
\newcommand{\bprob}{\begin{prob}}
\newcommand{\eprob}{\end{prob}}

\newcommand{\bvtm}{\begin{verbatim}}
\newcommand{\bfig}{\begin{figure}}
\newcommand{\efig}{\end{figure}}
\newcommand{\bcen}{\begin{center}}
\newcommand{\ecen}{\end{center}}
\newcommand{\bfrm}{\begin{frame}}
\newcommand{\efrm}{\end{frame}}

\def\Bbb{\mathbb }

\def\real {{\Bbb R}}

\long\def\comment#1{}

%% file: TranIT11toArxiv.bbl
\begin{thebibliography}{10}
\providecommand{\url}[1]{#1}
\csname url@samestyle\endcsname
\providecommand{\newblock}{\relax}
\providecommand{\bibinfo}[2]{#2}
\providecommand{\BIBentrySTDinterwordspacing}{\spaceskip=0pt\relax}
\providecommand{\BIBentryALTinterwordstretchfactor}{4}
\providecommand{\BIBentryALTinterwordspacing}{\spaceskip=\fontdimen2\font plus
\BIBentryALTinterwordstretchfactor\fontdimen3\font minus
  \fontdimen4\font\relax}
\providecommand{\BIBforeignlanguage}[2]{{%
\expandafter\ifx\csname l@#1\endcsname\relax
\typeout{** WARNING: IEEEtran.bst: No hyphenation pattern has been}%
\typeout{** loaded for the language `#1'. Using the pattern for}%
\typeout{** the default language instead.}%
\else
\language=\csname l@#1\endcsname
\fi
#2}}
\providecommand{\BIBdecl}{\relax}
\BIBdecl

\bibitem{network_sim_book}
R.~M. Fujimoto, K.~S. Perumalla, and G.~F. Riley,
  \emph{\BIBforeignlanguage{English}{Network Simulation}}.\hskip 1em plus 0.5em
  minus 0.4em\relax Morgan \& Claypool Publishers, 2007.

\bibitem{paral_sim_1}
R.~Bagrodia, R.~Meyer, M.~Takai, Y.~A. Chen, X.~Zeng, J.~Martin, and H.~Y.
  Song, ``Parsec: a parallel simulation environment for complex systems,''
  \emph{Computer}, vol.~31, no.~10, pp. 77 --85, Oct. 1998.

\bibitem{paral_sim_2}
H.~Plesser, J.~Eppler, A.~Morrison, M.~Diesmann, and M.~O. Gewaltig,
  ``Efficient parallel simulation of large-scale neuronal networks on clusters
  of multiprocessor computers,'' in \emph{Euro-Par 2007 Parallel Processing}.

\bibitem{kushnerbook}
H.~J. Kushner, \emph{Approximation and Weak Convergence Methods for Random
  Processes, with Applications to Stochastic Systems Theory}.\hskip 1em plus
  0.5em minus 0.4em\relax Cambridge, MA: MIT Press, 1984.

\bibitem{RePEc:spr:finsto:v:9:y:2005:i:4:p:519-537}
\BIBentryALTinterwordspacing
R.~Norberg, ``Anomalous {PDE}s in {M}arkov chains: Domains of validity and
  numerical solutions,'' \emph{Finance and Stochastics}, vol.~9, no.~4, pp.
  519--537, October 2005. [Online]. Available:
  \url{http://ideas.repec.org/a/spr/finsto/v9y2005i4p519-537.html}
\BIBentrySTDinterwordspacing

\bibitem{darling-2008-5}
\BIBentryALTinterwordspacing
R.~W.~R. Darling and J.~R. Norris, ``Differential equation approximations for
  {M}arkov chains,'' \emph{Probability Surveys}, vol.~5, p.~37, 2008. [Online].
  Available: \url{doi:10.1214/07-PS121}
\BIBentrySTDinterwordspacing

\bibitem{chongcontinuum}
E.~K.~P. Chong, D.~Estep, and J.~Hannig, ``Continuum modeling of large
  networks,'' \emph{Int. J. Numer. Model.}, vol.~21, no.~3, pp. 169--186, 2008.

\bibitem{PDE1}
S.~L. Sobolev, \emph{Partial Differential Equations of Mathematical
  Physics}.\hskip 1em plus 0.5em minus 0.4em\relax Courier Dover Publications,
  1964.

\bibitem{PDE2}
R.~G. Mortimer, \emph{Mathematics for Physical Chemistry}.\hskip 1em plus 0.5em
  minus 0.4em\relax Academic Press, 2005.

\bibitem{PDE3}
M.~Gillman, \emph{An Introduction to Mathematical Models in Ecology and
  Evolution: Time and Space}.\hskip 1em plus 0.5em minus 0.4em\relax
  Wiley-Blackwell, 2009.

\bibitem{PDE4}
T.~Hens and M.~O. Rieger, \emph{Financial Economics}.\hskip 1em plus 0.5em
  minus 0.4em\relax Springer, 2010.

\bibitem{FEM}
G.~R. Liu and S.~S. Quek, \emph{The Finite Element Method: A Practical
  Course}.\hskip 1em plus 0.5em minus 0.4em\relax Butterworth-Heinemann, 2003.

\bibitem{FDM}
A.~R. Mitchell and D.~F. Griffiths, \emph{The Finite Difference Method in
  Partial Differential Equations}.\hskip 1em plus 0.5em minus 0.4em\relax
  Wiley, 1980.

\bibitem{cont_mdl_phy_book}
E.~W. C.~v. Groesen and J.~Molenaar,
  \emph{\BIBforeignlanguage{English}{Continuum Modeling in the Physical
  Sciences}}.\hskip 1em plus 0.5em minus 0.4em\relax Society for Industrial and
  Applied Mathematics, 2007.

\bibitem{cont_mdl_mtrl_book}
H.~B. M\"{u}hlhaus, \emph{\BIBforeignlanguage{English}{Continuum Models for
  Materials with Microstructure}}.\hskip 1em plus 0.5em minus 0.4em\relax
  Wiley, 1995.

\bibitem{cont_mdl_traf_book}
W.~F. Phillips, \emph{\BIBforeignlanguage{English}{A New Continuum Model for
  Traffic Flow}}.\hskip 1em plus 0.5em minus 0.4em\relax U.S. Dept. of
  Transportation, Research and Special Programs Administration National
  Technical Information Service [distributor], 1981.

\bibitem{cont_mdl_bio_paper}
D.~Gr\"{u}nbaum, ``Translating stochastic density-dependent individual behavior
  with sensory constraints to an {E}ulerian model of animal swarming,'' \emph{J
  Math Biol}, vol.~33, pp. 139--161, 1994.

\bibitem{hvy_trffc_98}
\BIBentryALTinterwordspacing
J.~M. Harrison, ``\BIBforeignlanguage{English}{Heavy traffic analysis of a
  system with parallel servers: Asymptotic optimality of discrete-review
  policies},'' \emph{\BIBforeignlanguage{English}{The Annals of Applied
  Probability}}, vol.~8, no.~3, pp. pp. 822--848, 1998. [Online]. Available:
  \url{http://www.jstor.org/stable/2667208}
\BIBentrySTDinterwordspacing

\bibitem{hvy_trffc_09}
J.~G. Dai and J.~M. {Harrison}, ``Reflecting brownian motion in three
  dimensions: A new proof of sufficient conditions for positive recurrence,''
  \emph{Mathematical Methods of Operations Research}, 2009.

\bibitem{hvy_trffc_10}
M.~{Bramson}, J.~G. {Dai}, and J.~M. {Harrison}, ``{Positive recurrence of
  reflecting Brownian motion in three dimensions},'' \emph{ArXiv e-prints},
  Sep. 2010.

\bibitem{kumarnetworkcapacity}
P.~Gupta and P.~R. Kumar, ``The capacity of wireless networks,''
  \emph{Information Theory, IEEE Transactions on}, vol.~46, no.~2, pp.
  388--404, Mar 2000.

\bibitem{homo_cont_mdl}
E.~W. Grundke and A.~N.~Z. Heywood, ``A uniform continuum model for scaling of
  ad hoc networks,'' in \emph{ADHOC-NOW}, 2003, pp. 96--103.

\bibitem{mean_fld_paper1}
\BIBentryALTinterwordspacing
R.~Bakhshi, L.~Cloth, W.~Fokkink, and B.~R. Haverkort, ``Meanfield analysis for
  the evaluation of gossip protocols,'' \emph{SIGMETRICS Perform. Eval. Rev.},
  vol.~36, pp. 31--39, November 2008. [Online]. Available:
  \url{http://doi.acm.org/10.1145/1481506.1481513}
\BIBentrySTDinterwordspacing

\bibitem{mean_fld_paper2}
M.~E.~J. {Newman}, C.~{Moore}, and D.~J. {Watts}, ``Mean-field solution of the
  small-world network model,'' \emph{Physical Review Letters}, vol.~84, pp.
  3201--3204, Apr. 2000.

\bibitem{pdetextbook}
R.~B. Guenther and J.~W. Lee, \emph{Partial Differential Equations of
  Mathematical Physics and Integral Equations}.\hskip 1em plus 0.5em minus
  0.4em\relax Mineola, NY: Courier Dover Publications, 1996.

\bibitem{Gard_SDE}
T.~C. Gard, \emph{Introduction to Stochastic Differential Equations (Pure and
  Applied Mathematics)}.\hskip 1em plus 0.5em minus 0.4em\relax Marcel Dekker
  Inc, 1987.

\bibitem{kushner_yin}
H.~J. Kushner and G.~G. Yin, \emph{Stochastic Approximation and Recursive
  Algorithms and Applications}.\hskip 1em plus 0.5em minus 0.4em\relax
  Springer, 2003.

\bibitem{statbook}
P.~Billingsley, \emph{Probability and Measure}.\hskip 1em plus 0.5em minus
  0.4em\relax New York, NY: Wiley, 1995.

\end{thebibliography}
